\documentclass[journal]{IEEEtran}
\usepackage[pdftex]{graphicx}
\usepackage[cmex10]{amsmath}
\usepackage{amssymb}
\usepackage{booktabs}
\usepackage{caption}
\usepackage{subcaption}
\usepackage{cite}
\usepackage{psfrag}
\usepackage{epstopdf}
\usepackage{algorithm}
\usepackage{algorithmic}
\usepackage{widetext}
\usepackage{float}
\usepackage{color}
\usepackage{soul}
\usepackage{upgreek}
\usepackage{multirow}
\usepackage[justification=centerlast]{caption}
\usepackage{tikz,pgfplots}
\usepackage{dblfloatfix}
\usepackage{adjustbox}
\DeclareUnicodeCharacter{2005}{\hspace{0.01em}}
\usepackage{tikz} 
\usetikzlibrary{calc}
\usepackage{mathtools}
\usetikzlibrary{arrows}
\pgfplotsset{compat=newest}

\usetikzlibrary{patterns}
\usepgfplotslibrary{fillbetween}

\usepackage[pdftex]{graphicx}
\usepackage[cmex10]{amsmath}
\usepackage{amssymb}
\usepackage{caption}
\usepackage{subcaption}
\usepackage{cite}
\usepackage{psfrag}
\usepackage{epstopdf}
\usepackage{algorithm}
\usepackage{algorithmic}
\usepackage{widetext}
\usepackage{float}
\usepackage{color}
\usepackage{soul}
\usepackage{upgreek}
\usepackage{multirow}
\usepackage[justification=centerlast]{caption}
\usepackage{tikz,pgfplots,filecontents}
\usepackage{dblfloatfix}
\usepackage[numbers,sort&compress,square]{natbib}

\usepackage{tikz} 
\usetikzlibrary{calc}
\usepackage{mathtools}
\usetikzlibrary{arrows}
\usepackage{tikz,pgfplots,filecontents}
\pgfplotsset{compat=newest}
\usepgfplotslibrary{fillbetween}
\definecolor{myBlue}{RGB}{72,125,215}
\definecolor{myOrange}{RGB}{118,54,45}
\definecolor{InfinBlue}{RGB}{72,72,51}

\ifCLASSINFOpdf
\else
\fi
\begin{document}
%
\title{Transfer Learning for Neural Networks-based Equalizers in Coherent Optical Systems}
     \pgfplotsset{
        compat=1.3, 
        my axis style/.style={
            every axis plot post/.style={/pgf/number format/fixed},
            ybar=5pt,
            bar width=8pt,
            x=1.7cm,
            axis on top,
            enlarge x limits=0.1,
            symbolic x coords={MLP, biLSTM, ESN, CNN+MLP, CNN+biLSTM, DBP},
            visualization depends on=rawy\as\rawy, 
            nodes near coords={%
                \pgfmathprintnumber[precision=2]{\rawy}
            },
            every node near coord/.append style={rotate=90, anchor=west},
            tick label style={font=\footnotesize},
            xtick distance=1,
        },
    }
%

\author{Pedro J. Freire, Daniel Abode,
Jaroslaw E. Prilepsky, Nelson Costa\\ Bernhard Spinnler,
Antonio Napoli, Sergei K. Turitsyn
\thanks{This paper was supported by the EU Horizon 2020 program under the Marie Sklodowska-Curie grant agreement 813144 (REAL-NET). YO acknowledges the support of the SMARTNET EMJMD programme (Project number - 586686-EPP-1-2017-1-UK-EPPKA1-JMD-MOB). JEP is supported by Leverhulme Trust, Grant No. RP-2018-063. SKT acknowledges support of the EPSRC project TRANSNET}
\thanks{Pedro J. Freire, Daniel Abode, Jaroslaw E. Prilepsky and Sergei K. Turitsyn are with Aston Institute of Photonic Technologies, Aston University, United Kingdom, p.freiredecarvalhosouza@aston.ac.uk.}
\thanks{Antonio Napoli is with Infinera London, UK, anapoli@infinera.com.}
\thanks{Bernhard Spinnler is with Infinera R\&D, Sankt-Martin-Str. 76, 81541, Munich, Germany, bspinnler@infinera.com.}
\thanks{Nelson Costa is with Infinera Unipessoal, Lda, Rua da Garagem nº1, 2790-078 Carnaxide, Portugal, ncosta@infinera.com.}
\thanks{Manuscript received xxx 19, zzz; revised January 11, yyy.}}

%
%

\markboth{Journal of Lightwave technology , ~Vol.~y, No.~x, June~2021}%
{Shell \MakeLowercase{\textit{et al.}}: Transfer Learning for Neural Networks-based Equalizers in Coherent Optical Systems}
%



\maketitle
\begin{abstract}
In this work, we address the question of the adaptability of artificial neural networks (NNs) used for impairments mitigation in optical transmission systems. We demonstrate that by using well-developed techniques based on the concept of transfer learning, we can efficaciously retrain NN-based equalizers to adapt to the changes in the transmission system, using just a fraction (down to 1\%) of the initial training data or epochs. We evaluate the capability of transfer learning to adapt the NN to changes in the launch power, modulation format, symbol rate, or even fiber plants (different fiber types and lengths). The numerical examples utilize the recently introduced NN equalizer consisting of a convolutional layer coupled with bi-directional long-short term memory (biLSTM) recurrent NN element. Our analysis focuses on long-haul coherent optical transmission systems for two types of fibers: the standard single-mode fiber (SSMF) and the TrueWave Classic (TWC) fiber. We underline the specific peculiarities that occur when transferring the learning in coherent optical communication systems and draw the limits for the transfer learning efficiency. Our results demonstrate the effectiveness of transfer learning for the fast adaptation of NN architectures to different transmission regimes and scenarios, paving the way for engineering flexible and universal solutions for nonlinearity mitigation.  
\end{abstract}


%
\IEEEpeerreviewmaketitle

\section{Introduction}

\IEEEPARstart{T}{he} skyrocketing demand for capacity in optical communication systems continuously drives the search for efficient solutions to mitigate the impact of factors that degrade the performance of the optical line. In a long-haul optical network, the nonlinear effects are among the main degrading factors. Neural network (NN) methods have recently shown their efficiency for nonlinear compensation and noticeably improved transmission performance~\cite{Zibar19,zibar2020,hager2018nonlinear,khan2017machine,alan2020} compared to other solutions~\cite{Cart2017,dar2017,lin2014adaptive,bakhshali2015frequency}. Several NN-based approaches have been proposed, and we can roughly categorize them as either the ``black box'' models or the ``physics-inspired'' models \cite{hager2020}. The ``black box'' models are the NN architectures developed in other fields and reused to equalize the optical channels after training with the dataset from simulated or experimental optical transmission setups~\cite{freire2021performance,sidelnikov2018equalization,deligiannidis2021performance}. On the other hand, the ``physics-inspired'' NN architecture makes use of the known properties of channel models~\cite{freire2020complex, zhang2019field} or mimics the split-step-based digital back-propagation (DBP) technique~\cite{hager2018nonlinear, ghazisaeidi2020deep, Sidelnikov2021,Zibar19,zibar2020}. In the current work, to exemplify our method we deal with the black-box equalizer proposed in \cite{freire2021performance}.

We note that, despite the broad number of solutions and models considered in the literature, none of the studied NN-based equalizers has been converted into a practical working DSP solution so far. To become a practical impairment mitigation method, the NN-based equalizers must satisfy some straightforward requirements. First, it is natural to expect that the NN-based equalizers must perform better than the existing alternative methods, such as the Volterra or DBP equalizers~\cite{Cart2017,dar2017}. Also, the NN-equalizers' computational complexity should at least be comparable to that of other available DSP blocks~\cite{Cart2017}. Lastly, the NN-based solutions must be generalizable, i.e. they should demonstrate sufficient flexibility to work satisfactorily in different transmission scenarios with high adaptability and reconfigurability. This paper addresses the latter issue.

To ensure good performance, NN's training and test datasets must be independent and identically distributed.~\cite{tan2018survey}. This requirement may be hard to achieve in a commercial implementation since realistic optical network parameters are often dynamic. This means that we have to refit the equalizer for every change in the transmit power, modulation format, symbol rate, transmission reach, and the likes. Otherwise, the test and training data could be ``un-identically'' distributed. This limitation can cause a noticeable deterioration in the performance of an NN-based equalizer, as will be discussed further in this paper. One solution would be to train a new NN equalizer from scratch for each variation in channel and signal properties. However, such an approach is rather impractical and computationally inefficient, as a new training procedure can be prohibitively resource-hungry. Typically, an NN training procedure requires a considerable amount of training data and, often, a lot of training time to reach the best equalizer's performance under the new conditions. Thus, we need to identify alternative approaches to improve the flexibility and universality of the NN-based equalizers.

In our work, we propose employing a more effective approach that consists of using the well-developed machine learning technique -- the \textit{transfer learning} (TL) \cite{pan2009survey,tan2018survey,lu2015transfer,zhuang2020comprehensive}. Within the TL technique, we reuse some parameters from the NN model trained for the initial system (the \textit{source} domain) to build the new NN variation that fits the modified system (the \textit{target} domain), using a smaller amount of training resources. 

\begin{figure}[ht!]
\centering\includegraphics[width=0.49\textwidth]{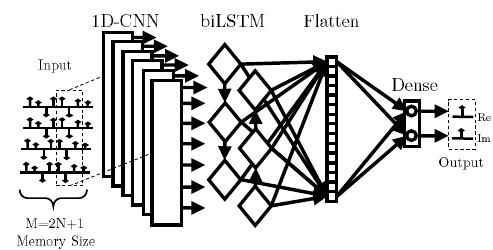}
\caption{Schematics of NN-based equalizer (CNN+biLSTM) used in this paper: the input consists of $M$ real and imaginary parts of the symbols for two polarizations where $N$ is the number of neighboring symbols considered by LSTM, while the output is composed of real and imaginary parts of the desired symbol. The input is sequentially processed by the convolutional layer, biLSTM layer, and then reshaped by the flattening and output layers. The NN is set to solve the regression task for real and imaginary parts of the desired symbol. For more details and equalizer's functionality metrics see~\cite[Section II, Fig. 3]{freire2021performance}.}
\label{fig:MODEL}
\end{figure}

In this paper, we present the methodology to transfer the learned features for a nonlinear optical channel across different launch powers, modulation formats (16-QAM to 32/64/128-QAM), symbol rate (34.4~GBd to 45/65/85~GBd), and fiber plant (9$\times$50~km TWC fiber to 18$\times$50~km SSMF). To demonstrate the functionality of the new technique, we use the recently-published efficient ``CNN+biLSTM'' equalizer ~\cite{freire2021performance}. It contains one convolutional layer coupled with one bidirectional LSTM layer as shown in Fig.~\ref{fig:MODEL}. However, we argue that the TL technique can be used with other NN-based equalizer architectures, as observed in other machine learning applications~\cite{tan2018survey,lu2015transfer}. The results obtained show a reduction in the number of epochs by up to 99\%, or (and) up to 99\% reduction in the size of the training dataset when employing the TL. This finding reveals the possibility of realizing a fast re-configurable nonlinear equalizer, thereby reducing the gap of implementing a practical NN-based nonlinear equalization for the next generation of optical networks. 

The remainder of the paper is organized as follows. The application of TL in optical communication systems is discussed in Sec.~\ref{sec:tranfer}. We also discuss the theoretical underpinnings of how this approach can be used to successfully equalize coherent optical channels. The simulator configuration, results, and discussions on the TL implementation are presented in Sec.~\ref{Sec:Result}. This includes a comparison of the cases: i) when the NN is trained with the TL; ii) when the NN is trained from scratch (random initialization of network parameters); iii) when the source NN is tested with a target dataset without retraining; iv) the reference case of employing linear equalization only. In the end, we discuss some limitations of the TL in the optical channel equalization problem. We conclude the paper with a summary of our approach and results.

\section{Transfer learning in Optical Fiber Communications}\label{sec:tranfer}
\subsection{Previous applications of transfer learning in optical fiber communications}\label{subsec:review}

The TL can be defined as a system's ability to identify and adapt knowledge acquired in previous tasks to new tasks~\cite{rosenstein2005transfer}.
Recent publications on the application of TL in optical communication focus on optical network tools\cite{musumeci2020transfer, cheng2020transfer, mo2018ann, xia2019transfer, gao2020ann, yao2019transductive}. A few works also addressed the nonlinearity mitigation issue~\cite{zhang2019fast, xu2020feedforward, zhang2020nonlinear}.

The TL in optical networks has been mainly used for optical signal-to-noise ratio (OSNR) monitoring. In~\cite{mo2018ann}, this application was introduced using an artificial NN-based TL approach to accurately predict the quality of the transmission of different optical networks without re-training NN models from scratch. In that paper, the source domain was a 4$\times$80~km (4 spans) large effective area fiber (LEAF) link using QPSK modulation. The target domain was the same system but with a different number of spans (propagation distance) and different modulations formats (4$\times$80~km LEAF with 16-QAM; 2$\times$80~km LEAF with 16-QAM; and 3$\times$80~km dispersion-shifted fiber with QPSK). The results showed that when using the TL, just 2\% of the original training dataset size was enough to calibrate the NN for the new target domain. More recently, in~Ref.~\cite{cheng2020transfer}, the experimental demonstration of the application of TL for joint OSNR monitoring and
modulation format identification from 64-QAM signals was presented. It was shown that by implementing the TL from simulation
to experiment, the number of training samples and epochs needed for the same prediction quality was reduced by 24.5\% and 44.4\%, respectively. 
Another recent application of TL was in the spectrum optimization problem for the resource reservation~\cite{yao2019transductive}. To predict a spectrum defragmentation time, the pre-trained NN model for a source domain (having a 6-node topology) was transferred and trained again using the data from the target domain (the NSFNet with 14 nodes). It was shown that by using the TL technique, the proportion of affected services was reduced, the overall likelihood of resource reservation failure was diminished, and the spectrum resource utilization was improved.

Only a few works have addressed the topic of TL for nonlinearity mitigation, and they focus on short-haul direct detection systems. In~\cite{xu2020feedforward} the successful transfer of the knowledge for the links with different bit rates and fiber lengths was demonstrated. Both feedforward and recurrent NNs were tested for the TL application: about 90\% (feed-forward) and 87.5\% (recurrent) reduction in epochs were achieved, and 62.5\% (feed-forward) and 53.8\% (recurrent) reduction in training symbols were demonstrated. Another work in direct detection, Ref.~\cite{zhang2019fast}, applied the TL from 5~dBm launch power to other powers (ranging from -7~dBm to 9~dBm) and from one transmission distance (640~km) to other ones (from 80~km up to 800~km). The experimental results showed that the iterations of TL constitute approximately one-fourth of the full NN training iterations. Additionally, the TL did not result in a performance penalty in a five-channel transmission when transferring the learned features from training just the middle channel to the four other channels. 

Finally, to the best of our knowledge, the TL in coherent optical systems was investigated in the only paper~\cite{zhang2020nonlinear}. In that work, the authors applied the TL for different launch powers but provided a very brief explanation of the technique. We would like to stress that compared to the previous works, in this paper we explicitly explain how we can successfully transfer the learned features from nonlinear optical channels using the NN equalizers, addressing the coherent transmission systems. We present a novel and broad description of how the TL can be efficiently used to realize flexible NN equalizers for adaptation to changes in launch power, modulation format, symbol rate, and fiber setup.

\subsection{Application of transfer learning to nonlinearity mitigation}\label{subsec:TransferL}
First, we identify the \textit{domain} and the \textit{task} notions for optical channel equalization. The domain consists of a feature space $Y$ which is an array of time-domain vectors (the memory window vectors). Each window vector consists of the real and imaginary parts of the received symbol (in each polarization) at time-step $k$ and its $2N$ neighboring (past and future) symbols. The task consists of two parts: i) the label space $X$, which is the set of real and imaginary parts of the transmitted symbols, and ii) the channel posterior function  $f$, which defines the conditional probabilistic distribution $p(X|Y)$. Several machine learning models can be used to learn the function $f$. Herein, we use the recently proposed NN-based equalizer "CNN+biLSTM"; for particular details see~\cite[Section II, Fig. 3]{freire2021performance} and Fig.~\ref{fig:MODEL}. We chose this architecture because it delivers the best performance for impairments mitigation in long-haul coherent optical systems when compared to several alternative NN structures, provided that the computational complexity is not restricted~\cite{freire2021performance, kotlyar2021}.

Having identified the domain and task in the sense of optical channel equalization, we now explain ``what to transfer'', i.e., which information can be transferred between different domains and tasks. The pass-averaged Manakov equation~\cite{gaiarin2018dual} describes the averaged evolution of the slowly varying complex-valued envelopes of the electric field in an optical fiber. From the Manakov equation, we can state that the channel likelihood, or the conditional probability of the received signal given the transmitted signal, $p(Y|X)$, is affected by the launch power, symbol rate, fiber type characterized by its attenuation coefficient $\alpha$, the dispersion coefficient $\beta_2$, the Kerr nonlinear coefficient $\gamma$, and the link setup, characterized, e.g., by the number of spans $N_s$, the span length $L$, the amplifier noise figure $NF$. In a nutshell, the NN-model that learns $f$ seeks to grasp the inverse correlation between the transmitted and received symbols, hence the aforementioned parameters are also important for determining our posterior function $f$.

As far as the propagation within both Task A (the source) and Task B (the target) is governed by the Manakov equation, we can apply the TL to boost the training performance of the Task B model from the Task A model. The TL strategy that fits this goal is the \textit{inductive TL}~\cite{tan2018survey}. In the inductive TL, the source and target tasks are different but related, and the TL seeks to strengthen the target task by exploiting the source domain's inductive biases. The transductive and unsupervised TLs are the two other TL techniques~\cite{zhuang2020comprehensive}. The definition of the unsupervised TL is analogous to inductive TL; however, since no labels are available in this case, we are unable to solve our purposed regression problem of channel equalization using this strategy. In the case of transductive TL, the source and target tasks must be identical but, as mentioned before, in our case, the task will change as the transmission parameters change. As a result of these constraints, transductive and unsupervised TLs are inapplicable to our current goal of nonlinearity mitigation in optical channel equalization.

From the model perspective, our TL uses the parameter control strategy~\cite{zhuang2020comprehensive}. In this case, the attribute priors, or probabilistic distribution parameters of the signal features, can be learned from the source domain and then be used to ease the learning of the target equalizer model. The parameters of a model reflect the knowledge learned by the model. As a consequence, the knowledge can be passed by sharing the parameters between different task models.

\begin{figure*}[ht!]
\centering\includegraphics[width=0.7\textwidth]{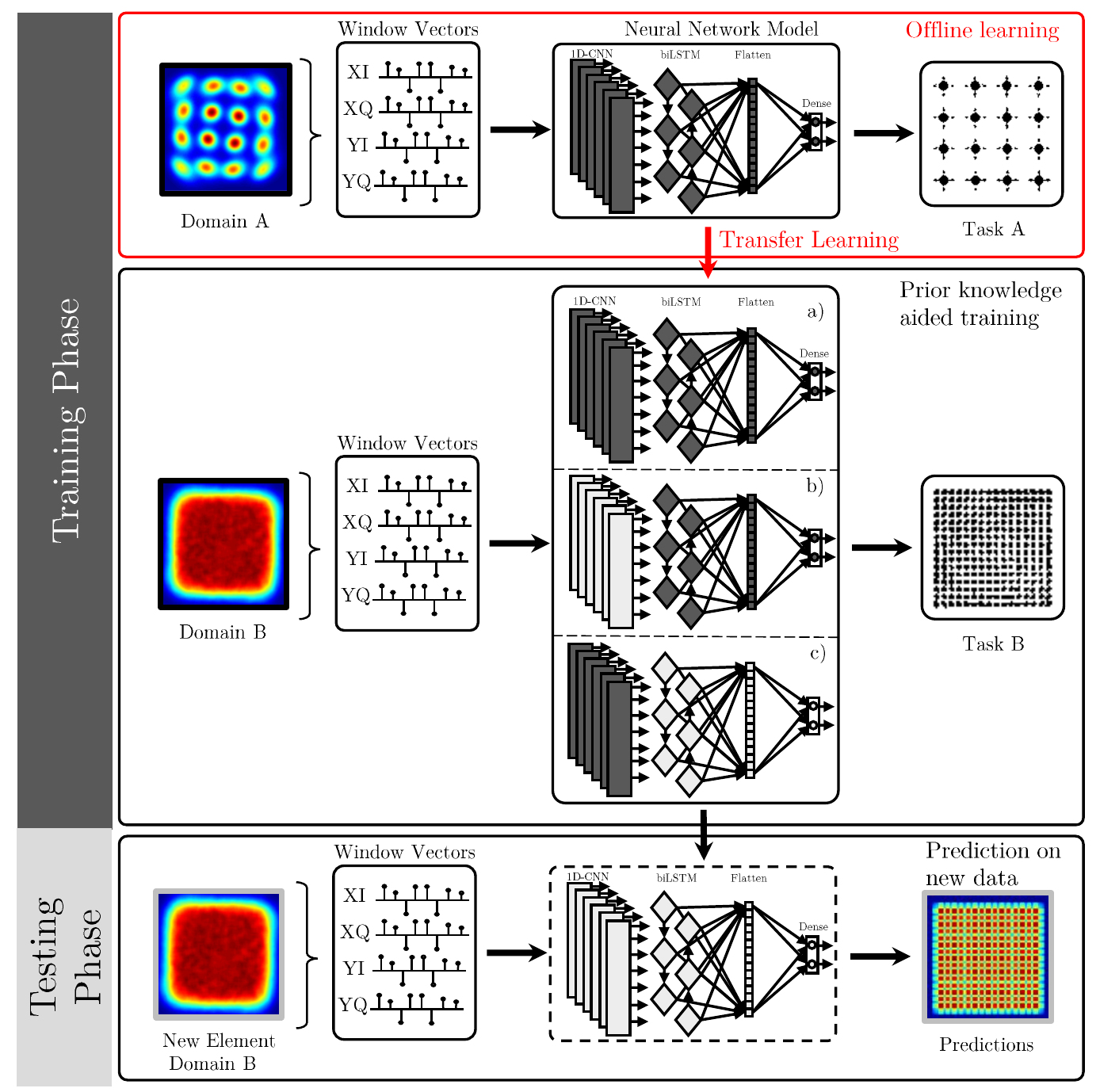}
\caption{Schematics of transferring the learning for the optical communication system with the NN-based equalizer (CNN+biLSTM). The leftmost subfigures display the received constellations, while the equalized constellations are shown in the rightmost ones. XI, XQ, YI, YQ refer to the I and Q components of X and Y polarizations. The NN elements that are getting trained/retrained are marked with dark gray, while the fixed NN elements are highlighted with light gray. The top panel represents the offline learning to train the NN model to a certain Domain/Task A. Three possible strategies of TL for the new Domain/Task B are shown in the middle panel: a) the model is retrained completely to the new Task B with the initial weights coming from the model trained on Task A; b) only the biLSTM and output layers are retrained to the new Task B and the convolutional layer is frozen with the weights coming from Task A; c) only the CNN layer is retrained to the new Task B and the biLSTM and output layers are frozen with the weights coming from Task A. In the lowest panel, we evaluate the performance of the NN with a completely new dataset for the Domain/Task B, keeping all layers frozen. }
\label{fig:TRANSFERLEARNING}
\end{figure*}

We note that the interpretability of composite neural network structures is still, largely, a matter of debate in the machine learning community~\cite{fan2021interpretability,zhang2018interpretable,zhang2018visual}. For our NN-based equalizer (CNN+biLSTM)~\cite{freire2021performance}, both the LSTM layer and the CNN layer jointly contribute to the nonlinear mitigation, but they do this in a complementary manner under their respective ``strengths''. However, some insights can be readily earned when analyzing the TL technique application to the considered CNN+biLSTM equalizer. In this paper, to save the training complexity during the transfer process, we did a standard procedure to test which layers are due to be retrained and which ones can be kept frozen. During our tests, we observed that whether we have to freeze/retrain a layer depends on the characteristics of the source and target domains. The implemented TL procedure is summarized in Fig. \ref{fig:TRANSFERLEARNING}. First, we train all the layers in the model using Domain/Task A (the source, top panel). We marked the NN equalizer layers that were (re)trained with a dark gray color, and the layers with the fixed weights with a light gray color. Next, we transfer the learned weights to the model for a new Domain/Task B (the target): three possible procedures for the transfer can be executed, as shown in the middle panel of Fig.~\ref{fig:TRANSFERLEARNING}. In the first case (case (a), the top inset in the middle panel), all the weights are learned again for the new Task B. Such an approach is recommended when there are considerable changes in nonlinearity and dispersion simultaneously (e.g., when we have a change in both the symbol rate and power). In the second case marked with (b), the middle part of the inset in the middle panel, the convolutional layer was frozen, and only the weights in the biLSTM and output layers are trainable. We can do this type of transfer (without losing the performance) to obtain a reduction in training complexity, when the channel memory changes noticeably, but the nonlinearity is still similar for both Tasks A and B. We have such a scenario when, for instance, we increase or decrease the symbol rate for Task B, but keep the same optical launch power for both Tasks A and B. Finally, in the third case (c) -- the lower part of the middle inset, -- the convolutional layer is trainable, but the biLSTM and output layers are frozen. This strategy evidently reduces the training complexity as well. This TL type can be used when the memory of the system is similar for both Tasks (e.g. the symbol rate is kept the same), but the nonlinearity for Task B changes, e.g. when we change the launch power. Finally, when we evaluate the performance of the new model attributed to Task B, we freeze all weights: this case is indicated in the lower panel of Fig.~\ref{fig:TRANSFERLEARNING}, entitled as ``Testing Phase''.  With this in mind, it is now possible to unveil some physical interpretations/motivations for our choice of the layers to be retrained. The reason for retraining only the CNN layer when the power changes, is because the hidden weights connecting different cells in the LSTM layer should remain fixed since the correlation between symbols remains almost the same. On the other hand, the kernel weights of each filter in the CNN layer need to be updated to account for the difference in the intensity of the nonlinearity. A similar logic can be followed when the memory of the transmission system changes. In this case, only the LSTM needs to be retrained since the weights between cells must learn the new correlation of the domain while the CNN can be kept frozen. In the next section, we will analyze how efficiently it is to transfer the learned network parameters from Domain/Task A to Domain/Task B for different modifications in the transmission parameters. 


\section{Results and discussion}\label{Sec:Result}

\begin{figure*}[ht!]
\centering\includegraphics[width=0.7\textwidth]{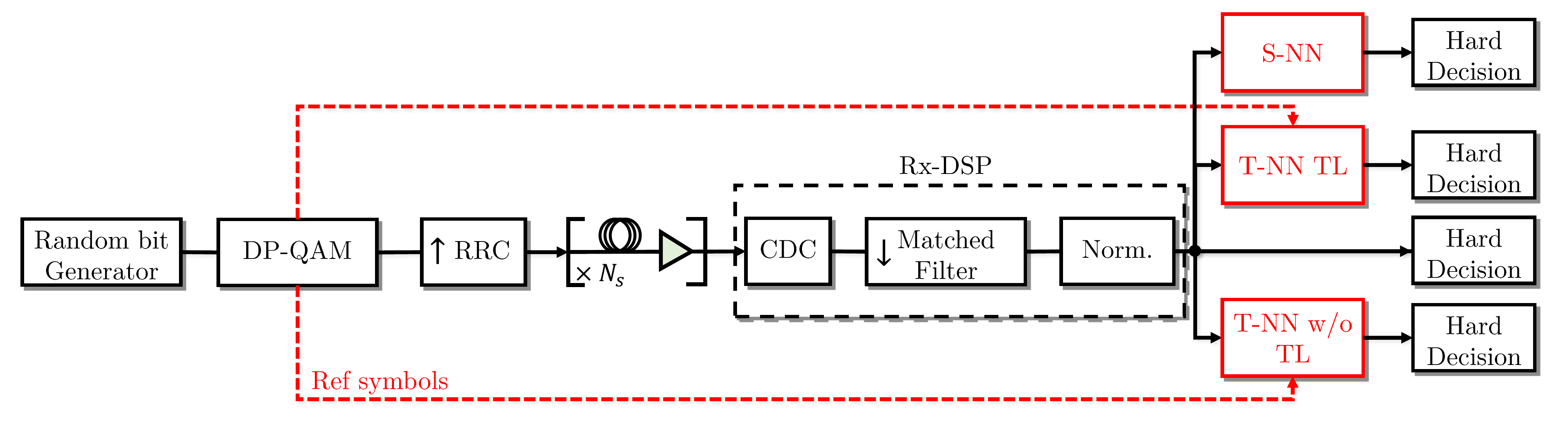}
\caption{Scheme of the numerical setup considered in our work, where the red elements indicate the 3 possible different NN implementations evaluated. The red arrows show that we use the transmitted symbols for the regression retraining but, since the S-NN is not retrained, it does not require receiving any transmitted symbols of the target domain. The explanations for each system's element are given in the main text, Subsec.~\ref{subSec:stup}.}
\label{fig:setup}
\end{figure*}

In this section, we explicitly evaluate the efficiency of the TL technique in terms of reducing the size of the training dataset and the number of epochs needed to reach acceptable accuracy of signal recovery. For comparison purposes, we use four key reference curves. i) only linear equalization is applied, where the respective Q-factor level is independent of the number of epochs, as no training occurs. This case is labeled as ``w/o NN'' and plotted with an orange straight line. The efficiency of the NN equalizer is measured against this curve. ii) The next reference curve is used to demonstrate the impact of changes in transmission parameters on the performance of the NN-based equalizer. In this case, the purple curve (labeled as the source NN, S-NN) shows the Q-factor when the NN equalizer is trained with the source Domain/Task A data only and tested on the target Domain/Task B without retraining. iii) We also evaluate the impact of training the NN using the data from the target Domain/Task B without using TL. This curve is labeled as ``T-NN w/o TL'', (T-NN means the target NN). In this case, the weights are initialized randomly, which corresponds to the traditional training of the NN equalizers. iv) Finally, we include the approach proposed in this paper, corresponding to transferring the learned parameters from the source Domain/Task A. We denote this as ``T-NN TL x \%'', where the ``x'' value represents the percentage of data used to train when compared to the T-NN w/o TL.

The following subsections address the details of the simulation setup and how the NN was trained. The results of the TL will also be shown for the changes in fiber type (using the TL of Fig.~\ref{fig:TRANSFERLEARNING}~(a)), symbol rate (using the TL of Fig.~\ref{fig:TRANSFERLEARNING}~(b)), and launch power and modulation format (using the TL of Fig.~\ref{fig:TRANSFERLEARNING}~(c)). We end this section with a short discussion of the TL limitations for our task.

\subsection{Numerical Setup and Neural Network model}\label{subSec:stup}
 
To illustrate the effect of the application of the proposed TL to the NN-based optical channel equalizers, we numerically simulated the dual-polarization (DP) transmission of a single-channel signal at 34.4 / 45 / 65 / 85 GBd. The signal is pre-shaped with a root-raised cosine (RRC) filter with 0.1 roll-off at a sampling rate of 8 samples per symbol. In addition, the signal had four possible modulation formats: 16 / 32 / 64 / 128-QAM. We tested modulation formats up to 128QAM since recent publications showed performance analysis for these modulation formats using simulation as well as experiment setups~\cite{qamar2020128,sarwar2018performance,khanna2016single}. We considered the following two test cases: (i) transmission over an optical link consisting of $9\!\times\!50$~km TWC spans; and (ii) transmission over $18\!\times\!50$~km SSMF spans. The optical signal propagation along the fiber was simulated by solving the Manakov equation via split-step Fourier method (SSFM)~\cite{agrawal2013nonlinear} with the resolution of $1$ km per step. The considered parameters of the TWC fiber are: the attenuation parameter $\alpha = 0.23$ dB/km, the dispersion coefficient $D = 2.8$ ps/(nm$\cdot$km), and the effective nonlinearity coefficient $\gamma = 2.5$~(W$\cdot$km)$^{-1}$. The SSMF parameters are: $\alpha = 0.2$ dB/km, $D = 17$ ps/(nm$\cdot$km), and $\gamma = 1.2$ (W$\cdot$km)$^{-1}$. Every span was followed by an optical amplifier with the noise figure $\text{NF}=4.5$~dB, which fully compensated fiber losses and added amplified spontaneous emission (ASE) noise. At the receiver, a standard Rx-DSP was used. It includes the full electronic chromatic dispersion compensation (CDC) using a frequency-domain equalizer, the application of a matched filter, and the downsampling to the symbol rate. Finally, the received symbols were normalized (by phase and amplitude) to the transmitted ones. No other transceiver distortions were considered. 

After the Rx-DSP, we estimate the bit error rate (BER) using the transmitted symbols, received soft symbols, and hard decisions after equalization, addressing the four cases depicted on the right side of Fig.~\ref{fig:setup}. The NN input mini-batch shape can be defined by three dimensions \cite{freire2021performance}: $(B, M, 4)$. $B$ is the mini-batch size, $M$ is the memory size defined through the number of neighbors $N$ as $M = 2N + 1$, and $4$ is the number of features for each symbol, referring to the real and imaginary parts of two polarization components. The output target is to recover the real and imaginary parts of the $k$-th symbol in one of the polarization, so the shape of the NN output batch can be expressed as $(B,2)$.

\begin{figure*}[htb]
    \centering 
\begin{subfigure}{0.33\textwidth}
  \begin{tikzpicture}[scale=0.65]
    \begin{axis} [
        xlabel={Number of epochs},
        ylabel={Q-Factor [dB]},
        grid=both,          xtick = {1,20,40,60,80,100},   
    	xmin=1, xmax=100,
        legend style={legend style={ at={(1,0.3)},anchor= east}, legend cell align=left,fill=white, fill opacity=0.6, draw opacity=1,text opacity=1},
    	grid style={dashed}]
        ]
      \addplot[color=red, mark=square, very thick]
    coordinates {(1,8.6023)(2,8.6434)(3,8.6502)(5,8.6348)(10,8.6523)(15,8.6523)(20,8.6425)(25,8.6614)(30,8.6318)(35,8.6485)(41,8.6446)(45,8.6473)(50,8.638)(55,8.6499)(60,8.6443)(65,8.6523)(70,8.6312)(75,8.6404)(80,8.644)(85,8.6482)(90,8.6315)(95,8.644)(100,8.6538)};
    \addlegendentry{\footnotesize{T-NN TL 100\%}};

    \addplot[color=green, mark=diamond, very thick]
  coordinates {(1,7.588)(2,7.7644)(3,7.588)(5,7.5577)(10,8.1197)(15,8.2731)(20,8.1197)(25,8.3439)(30,8.32)(35,8.32)(41,8.2964)(45,8.2731)(50,8.2964)(55,8.2964)(60,8.2503)(65,8.1197)(70,8.2731)(75,8.2503)(80,8.2277)(85,8.2964)(90,8.2503)(95,8.2731)(100,8.2964)};
    \addlegendentry{\footnotesize{T-NN TL 1\%}};
    
    \addplot[color=blue, mark=o, very thick]   
    coordinates {(1,3.2651)(2,3.3075)(3,3.4553)(5,4.8267)(10,6.0424)(15,6.6736)(20,6.7448)(25,7.1025)(30,7.356)(35,7.686)(41,7.898)(45,7.9735)(50,8.0136)(55,8.1076)(60,8.2137)(65,8.3328)(70,8.2123)(75,8.3923)(80,8.4051)(85,8.4618)(90,8.5476)(95,8.5487)(100,8.5811)};
    \addlegendentry{\footnotesize{T-NN w/o TL}};
    
    \addplot[color=violet, mark=*, very thick]     
    coordinates {(1,7.3071)(2,7.3071)(3,7.3071)(5,7.3071)(10,7.3071)(15,7.3071)(20,7.3071)(25,7.3071)(30,7.3071)(35,7.3071)(41,7.3071)(45,7.3071)(50,7.3071)(55,7.3071)(60,7.3071)(65,7.3071)(70,7.3071)(75,7.3071)(80,7.3071)(85,7.3071)(90,7.3071)(95,7.3071)(100,7.3071)};
    \addlegendentry{\footnotesize{S-NN}};

    \addplot[color=orange, mark=triangle, very thick]     
    coordinates {(1,3.5511)(2,3.5511)(3,3.5511)(5,3.5511)(10,3.5511)(15,3.5511)(20,3.5511)(25,3.5511)(30,3.5511)(35,3.5511)(41,3.5511)(45,3.5511)(50,3.5511)(55,3.5511)(60,3.5511)(65,3.5511)(70,3.5511)(75,3.5511)(80,3.5511)(85,3.5511)(90,3.5511)(95,3.5511)(100,3.5511)};
    \addlegendentry{\footnotesize{w/o NN}};
  \end{axis}
    \end{tikzpicture}
  \caption{ }
\end{subfigure}\hfil 
\begin{subfigure}{0.33\textwidth}
  \begin{tikzpicture}[scale=0.65]
    \begin{axis} [
        xlabel={Number of epochs},
        ylabel={Q-Factor [dB]},
        grid=both,          xtick = {1,20,40,60,80,100},   
    	xmin=1, xmax=100,
        legend style={legend style={ at={(1,0.3)},anchor= east}, legend cell align=left,fill=white, fill opacity=0.6, draw opacity=1,text opacity=1},
    	grid style={dashed}]
        ]
      \addplot[color=red, mark=square, very thick]
    coordinates {(1,9.9393)(2,10.0569)(3,10.0842)(5,10.1503)(10,10.1758)(15,10.1817)(20,10.229)(25,10.2315)(30,10.1984)(35,10.2056)(41,10.2166)(45,10.239)(50,10.19)(55,10.2228)(60,10.1936)(65,10.1972)(70,10.1793)(75,10.2008)(80,10.2179)(85,10.1972)(90,10.2315)(95,10.2069)(100,10.2215)};
    \addlegendentry{\footnotesize{T-NN TL 100\%}};

    \addplot[color=green, mark=diamond, very thick]
  coordinates {(1,8.0665)(2,8.3345)(3,8.8492)(5,9.163)(10,9.6858)(15,9.9814)(20,9.9814)(25,9.9814)(30,10.0204)(35,10.0405)(41,10.0204)(45,9.9814)(50,10.148)(55,10.1711)(60,10.1711)(65,10.1948)(70,10.1948)(75,10.1711)(80,10.1948)(85,10.2697)(90,10.2191)(95,10.2191)(100,10.1711)};
    \addlegendentry{\footnotesize{T-NN TL 5\%}};
    
    \addplot[color=blue, mark=o, very thick]   
    coordinates {(1,4.9936)(2,5.0257)(3,5.0862)(5,6.1375)(10,7.9007)(15,8.2068)(20,8.4083)(25,8.625)(30,8.8344)(35,9.3647)(41,9.5291)(45,9.5997)(50,9.5782)(55,9.8107)(60,9.7628)(65,9.7774)(70,9.9824)(75,10.1222)(80,9.9448)(85,10.1458)(90,10.0831)(95,10.1607)(100,10.1134)};
    \addlegendentry{\footnotesize{T-NN w/o TL}};
    
    \addplot[color=violet, mark=*, very thick]     
    coordinates {(1,6.4075)(2,6.4075)(3,6.4075)(5,6.4075)(10,6.4075)(15,6.4075)(20,6.4075)(25,6.4075)(30,6.4075)(35,6.4075)(41,6.4075)(45,6.4075)(50,6.4075)(55,6.4075)(60,6.4075)(65,6.4075)(70,6.4075)(75,6.4075)(80,6.4075)(85,6.4075)(90,6.4075)(95,6.4075)(100,6.4075)};
    \addlegendentry{\footnotesize{S-NN}};

    \addplot[color=orange, mark=triangle, very thick]  
    coordinates {(1,5.2159)(2,5.2159)(3,5.2159)(5,5.2159)(10,5.2159)(15,5.2159)(20,5.2159)(25,5.2159)(30,5.2159)(35,5.2159)(41,5.2159)(45,5.2159)(50,5.2159)(55,5.2159)(60,5.2159)(65,5.2159)(70,5.2159)(75,5.2159)(80,5.2159)(85,5.2159)(90,5.2159)(95,5.2159)(100,5.2159)};

    \addlegendentry{\footnotesize{W/o NN}};
  \end{axis}
    \end{tikzpicture}
  \caption{ }
\end{subfigure}\hfil 
\begin{subfigure}{0.33\textwidth}
  \begin{tikzpicture}[scale=0.65]
    \begin{axis} [
        xlabel={Number of epochs},
        ylabel={Q-Factor [dB]},
        grid=both,          xtick = {1,20,40,60,80,100},   
    	xmin=1, xmax=100,
        legend style={legend style={ at={(1,0.3)},anchor= east}, legend cell align=left,fill=white, fill opacity=0.6, draw opacity=1,text opacity=1},
    	grid style={dashed}]
        ]
      \addplot[color=red, mark=square, very thick]
    coordinates {(1,10.6854)(2,11.0888)(3,11.2436)(5,11.4483)(10,11.5655)(15,11.6097)(20,11.633)(25,11.5872)(30,11.5946)(35,11.5241)(41,11.6174)(45,11.5514)(50,11.6097)(55,11.633)(60,11.5376)(65,11.5946)(70,11.633)(75,11.5946)(80,11.6174)(85,11.5872)(90,11.5655)(95,11.5308)(100,11.5727)};

    \addlegendentry{\footnotesize{T-NN TL 100\%}};
    
    \addplot[color= black, mark=star, very thick]
  coordinates {(1,9.7971)(2,10.6854)(3,11.0013)(4,11.1368)(5,11.2261)(10,11.4424)(12,11.4664)(21,11.5043)(23,11.5308)(26,11.5308)(34,11.5376)(41,11.5655)(42,11.5727)(47,11.5655)(52,11.5872)(57,11.5946)(61,11.5727)(69,11.5514)(72,11.5727)(77,11.5514)(83,11.5584)(90,11.5655)(94,11.5655)(98,11.5872)(101,11.5655)};
    \addlegendentry{\footnotesize{T-NN TL 50\%}};
    
    \addplot[color= brown, mark=x, very thick]
  coordinates {(1,8.6817)(2,9.403)(3,10.0477)(4,10.5227)(5,10.7289)(11,11.1923)(16,11.3749)(20,11.4365)(26,11.4788)(31,11.5108)(40,11.5584)(46,11.5799)(53,11.5727)(59,11.641)(64,11.691)(69,11.6174)(75,11.6572)(76,11.641)(83,11.6572)(90,11.6739)(96,11.6655)(97,11.6739)(102,11.641)};
    \addlegendentry{\footnotesize{T-NN TL 20\%}};
    
    \addplot[color=cyan, mark=pentagon, very thick]
  coordinates {(1,7.353)(2,8.6526)(3,9.0894)(4,9.341)(5,9.9257)(11,10.6361)(16,11.0437)(23,11.1522)(30,11.28)(36,11.3283)(40,11.4365)(45,11.3802)(47,11.3802)(53,11.4365)(61,11.641)(71,11.5655)(77,11.5655)(84,11.5655)(88,11.4978)(95,11.5655)(97,11.641)};
    \addlegendentry{\footnotesize{T-NN TL 10\%}};
    
    \addplot[color=green, mark=diamond, very thick]
  coordinates {(1,7.6599)(2,7.4022)(3,8.3249)(5,9.012)(10,9.8067)(15,10.3514)(20,10.7892)(25,11.0109)(30,11.0782)(35,11.1522)(41,11.2348)(45,11.0782)(50,11.3283)(55,11.2348)(60,11.2348)(65,11.4365)(70,11.1522)(75,11.2348)(80,11.2348)(85,11.3283)(90,11.2348)(95,11.2348)(100,11.2348)};
    \addlegendentry{\footnotesize{T-NN TL 5\%}};
    
    \addplot[color=yellow, mark=|, very thick]
  coordinates {(1,6.5571)(2,7.2787)(3,7.9049)(4,7.9049)(5,7.8509)(6,7.8509)(16,8.3931)(21,8.7946)(26,9.1809)(30,9.4329)(33,9.4329)(39,9.7595)(44,9.9255)(51,10.1254)(56,10.2439)(61,11.0108)(67,11.4364)(71,11.4364)(76,11.4364)(81,11.4364)(86,11.4364)(91,11.4364)(96,11.4364)(100,11.4364)};
    \addlegendentry{\footnotesize{T-NN TL 1\%}};
    
    \addplot[color=blue, mark=o, very thick]   
    coordinates {(1,6.602)(2,6.6422)(3,6.7003)(5,7.4735)(10,8.9203)(15,9.5597)(20,9.9025)(25,10.1618)(30,10.3716)(35,10.1793)(41,10.4558)(45,10.8493)(50,11.0173)(55,10.9981)(60,10.8976)(65,11.1106)(70,11.28)(75,11.4365)(80,11.4483)(85,11.5043)(90,11.5872)(95,11.5886)(100,11.5655)};
    \addlegendentry{\footnotesize{T-NN w/o TL}};
    
    \addplot[color=violet, mark=*, very thick]     
    coordinates {(1,5.2675)(2,5.2675)(3,5.2675)(5,5.2675)(10,5.2675)(15,5.2675)(20,5.2675)(25,5.2675)(30,5.2675)(35,5.2675)(41,5.2675)(45,5.2675)(50,5.2675)(55,5.2675)(60,5.2675)(65,5.2675)(70,5.2675)(75,5.2675)(80,5.2675)(85,5.2675)(90,5.2675)(95,5.2675)(100,5.2675)};
    \addlegendentry{\footnotesize{S-NN}};

    \addplot[color=orange, mark=triangle, very thick]  
    coordinates {(1,6.767)(2,6.767)(3,6.767)(5,6.767)(10,6.767)(15,6.767)(20,6.767)(25,6.767)(30,6.767)(35,6.767)(41,6.767)(45,6.767)(50,6.767)(55,6.767)(60,6.767)(65,6.767)(70,6.767)(75,6.767)(80,6.767)(85,6.767)(90,6.767)(95,6.767)(100,6.767)};
    \addlegendentry{\footnotesize{w/o NN}};
  \end{axis}
    \end{tikzpicture}
  \caption{ }
\end{subfigure}

\medskip
\begin{subfigure}{0.33\textwidth}
    \begin{tikzpicture}[scale=0.65]
    \begin{axis} [
        xlabel={Number of epochs},
        ylabel={Q-Factor [dB]},
        grid=both,          xtick = {1,20,40,60,80,100},   
    	xmin=1, xmax=100,
        legend style={legend style={ at={(1,0.3)},anchor= east}, legend cell align=left,fill=white, fill opacity=0.6, draw opacity=1,text opacity=1},
    	grid style={dashed}]
        ]
      \addplot[color=red, mark=square, very thick]
    coordinates {(1,12.7783)(2,12.5763)(3,12.6671)(5,12.5763)(10,12.7196)(15,12.8452)(20,12.6197)(25,12.7196)(30,12.7196)(35,12.4994)(41,12.6197)(45,12.6197)(50,12.4324)(55,12.6197)(60,12.6197)(65,12.6197)(70,12.4994)(75,12.7783)(80,12.4648)(85,12.7196)(90,12.6197)(95,12.6197)(100,12.6197)};
    \addlegendentry{\footnotesize{T-NN TL 100\%}};

     \addplot[color=green, mark=diamond, very thick]
     coordinates {(1,11.1106)(2,10.5297)(3,9.9078)(8,9.885)(10,10.571)(15,11.4424)(20,11.0404)(25,11.0471)(30,11.8591)(35,12.0931)(41,12.0602)(45,12.3729)(50,12.6197)(55,12.6197)(60,12.7783)(65,12.7783)(70,12.7196)(75,12.6197)(80,12.7196)(85,12.6671)(90,12.6671)(95,12.6671)(100,12.6671)};
     \addlegendentry{\footnotesize{T-NN TL 1\%}};

    \addplot[color=blue, mark=o, very thick]   
    coordinates {(1,3.1194)(2,3.3706)(3,5.6361)(5,7.9613)(10,9.8492)(15,10.3834)(20,10.7267)(25,10.7773)(30,11.0782)(35,11.172)(41,11.2048)(45,11.1561)(50,11.5445)(55,11.5043)(60,11.6097)(65,11.633)(70,11.7453)(75,11.9189)(80,11.8259)(85,11.7453)(90,11.8943)(95,12.1103)(100,12.0141)};
    \addlegendentry{\footnotesize{T-NN w/o TL}};
    
    \addplot[color=violet, mark=*, very thick]     
    coordinates {(1,10.8866)(2,10.8866)(3,10.8866)(5,10.8866)(10,10.8866)(15,10.8866)(20,10.8866)(25,10.8866)(30,10.8866)(35,10.8866)(41,10.8866)(45,10.8866)(50,10.8866)(55,10.8866)(60,10.8866)(65,10.8866)(70,10.8866)(75,10.8866)(80,10.8866)(85,10.8866)(90,10.8866)(95,10.8866)(100,10.8866)};

    \addlegendentry{\footnotesize{S-NN}};

    \addplot[color=orange, mark=triangle, very thick]     
    coordinates {(1,3.2621)(2,3.2621)(3,3.2621)(5,3.2621)(10,3.2621)(15,3.2621)(20,3.2621)(25,3.2621)(30,3.2621)(35,3.2621)(41,3.2621)(45,3.2621)(50,3.2621)(55,3.2621)(60,3.2621)(65,3.2621)(70,3.2621)(75,3.2621)(80,3.2621)(85,3.2621)(90,3.2621)(95,3.2621)(100,3.2621)};
    \addlegendentry{\footnotesize{w/o NN}};
  \end{axis}
    \end{tikzpicture}
  \caption{ }
\end{subfigure}\hfil 
\begin{subfigure}{0.33\textwidth}
   \begin{tikzpicture}[scale=0.65]
    \begin{axis} [
        xlabel={Number of epochs},
        ylabel={Q-Factor [dB]},
        grid=both,          xtick = {1,20,40,60,80,100},   
    	xmin=1, xmax=100,
        legend style={legend style={ at={(1,0.3)},anchor= east}, legend cell align=left,fill=white, fill opacity=0.6, draw opacity=1,text opacity=1},
    	grid style={dashed}]
        ]
      \addplot[color=red, mark=square, very thick]
    coordinates {(1,10.244)(2,11.633)(3,12.8452)(8,13.5576)(10,13.2952)(15,13.2952)(20,13.2952)(25,13.2952)(30,13.2952)(35,13.2952)(41,13.1342)(45,13.2952)(50,13.1342)(55,13.1342)(60,13.1342)(65,13.1342)(70,13.1342)(75,13.1342)(80,13.1342)(85,13.1342)(90,13.1342)(95,13.1342)(100,13.1342)};
    \addlegendentry{\footnotesize{T-NN TL 100\%}};

     \addplot[color=green, mark=diamond, very thick]
   coordinates {(1,7.8617)(2,7.5943)(3,8.6922)(8,9.8902)(10,10.3804)(15,11.0109)(20,12.2946)(25,12.2946)(30,12.2946)(35,12.2946)(41,12.2946)(45,12.2946)(50,12.2946)(55,12.2946)(60,12.2946)(65,12.2946)(70,12.2946)(75,12.2946)(80,12.2946)(85,12.2946)(90,12.2946)(95,12.2946)(100,12.2946)};
    \addlegendentry{\footnotesize{T-NN TL 5\%}};
    
    \addplot[color=blue, mark=o, very thick]   
    coordinates {(1,4.1448)(2,4.8356)(3,4.8974)(8,8.1866)(10,9.0432)(15,9.7971)(20,10.7539)(25,10.8921)(30,11.164)(35,11.691)(41,11.6655)(45,12.1653)(50,12.0291)(55,12.2262)(60,12.0764)(65,12.2262)(70,12.4324)(75,12.4018)(80,12.3456)(85,12.6197)(90,12.6197)(95,12.6197)(100,12.6671)};
    \addlegendentry{\footnotesize{T-NN w/o TL}};
    
    \addplot[color=violet, mark=*, very thick]     
    coordinates {(1,6.7637)(2,6.7637)(3,6.7637)(5,6.7637)(10,6.7637)(15,6.7637)(20,6.7637)(25,6.7637)(30,6.7637)(35,6.7637)(41,6.7637)(45,6.7637)(50,6.7637)(55,6.7637)(60,6.7637)(65,6.7637)(70,6.7637)(75,6.7637)(80,6.7637)(85,6.7637)(90,6.7637)(95,6.7637)(100,6.7637)};
    \addlegendentry{\footnotesize{S-NN}};

    \addplot[color=orange, mark=triangle, very thick]     
    coordinates {(1,4.9392)(2,4.9392)(3,4.9392)(8,4.9392)(10,4.9392)(15,4.9392)(20,4.9392)(25,4.9392)(30,4.9392)(35,4.9392)(41,4.9392)(45,4.9392)(50,4.9392)(55,4.9392)(60,4.9392)(65,4.9392)(70,4.9392)(75,4.9392)(80,4.9392)(85,4.9392)(90,4.9392)(95,4.9392)(100,4.9392)};
    \addlegendentry{\footnotesize{w/o NN}};
  \end{axis}
    \end{tikzpicture}
  \caption{ }
\end{subfigure}\hfil 
\begin{subfigure}{0.33\textwidth}
  \begin{tikzpicture}[scale=0.65]
    \begin{axis} [
        xlabel={Number of epochs},
        ylabel={Q-Factor [dB]},
        grid=both, 
        xtick = {1,20,40,60,80,100},
    	xmin=1, xmax=100,
        legend style={legend style={ at={(1,0.3)},anchor= east}, legend cell align=left,fill=white, fill opacity=0.6, draw opacity=1,text opacity=1},
    	grid style={dashed}]
        ]
      \addplot[color=red, mark=square, very thick]
    coordinates {(1,8.4086)(2,10.0842)(3,11.2348)(8,13.2952)(10,13.5576)(15,13.5576)(20,13.5576)(25,13.5576)(30,13.5576)(35,13.5576)(41,13.5576)(45,13.5576)(50,13.5576)(55,13.5576)(60,13.5576)(65,13.5576)(70,13.5576)(75,13.5576)(80,13.5576)(85,13.5576)(90,13.5576)(95,13.5576)(100,13.5576)};

    \addlegendentry{\footnotesize{T-NN TL 100\%}};
    
     \addplot[color= black, mark=star, very thick]
  coordinates {(1,8.2337)(2,10.224)(3,11.2603)(4,11.8262)(5,12.5576)(7,12.976)(11,13.276)(16,13.5576)(21,13.5576)(26,13.5576)(31,13.5576)(36,13.5576)(41,13.5576)(46,13.5576)(51,13.5576)(56,13.5576)(61,13.5576)(66,13.5576)(71,13.5576)(76,13.5576)(81,13.5576)(86,13.5576)(91,13.5576)(96,13.5576)(101,13.5576)};
    \addlegendentry{\footnotesize{T-NN TL 50\%}};
    
    \addplot[color=brown, mark=diamond, very thick]
   coordinates {(1,7.0118)(2,7.8653)(3,8.8579)(4,9.5937)(5,10.1255)(11,12.923)(16,13.123)(21,13.323)(26,13.423)(31,13.523)(36,13.5523)(41,13.5576)(45,13.5576)(50,13.5576)(55,13.5576)(60,13.5576)(65,13.5576)(70,13.5576)(75,13.5576)(80,13.5576)(85,13.5576)(90,13.5576)(95,13.5576)(100,13.5576)};
    \addlegendentry{\footnotesize{T-NN TL 20\%}};

    \addplot[color=cyan, mark=pentagon, very thick]
    coordinates {(1,6.1201)(2,6.9919)(3,7.5579)(4,7.809)(5,8.5655)(11,10.3233)(16,11.4978)(19,11.9446)(23,12.2946)(28,12.4946)(31,12.5946)(36,12.6946)(41,12.7946)(46,12.8946)(50,12.9576)(55,13.0576)(60,13.1576)(65,13.1976)(70,13.1876)(75,13.2576)(80,13.3576)(85,13.5576)(90,13.5576)(95,13.5576)(100,13.5576)};
    \addlegendentry{\footnotesize{T-NN TL 10\%}};

     \addplot[color=green, mark=diamond, very thick]
   coordinates {(1,5.7771)(2,6.3571)(3,6.2226)(8,7.3915)(10,7.7502)(15,8.341)(20,8.9762)(25,9.6169)(30,10.0477)(35,10.5349)(41,10.9403)(45,11.1143)(50,11.4726)(55,11.8706)(60,12.0764)(65,12.3729)(70,12.5364)(75,12.7783)(80,13.1342)(85,13.0165)(90,13.1342)(95,13.2952)(100,13.1342)};
     \addlegendentry{\footnotesize{T-NN  TL 5\%}};

     \addplot[color=yellow, mark=|, very thick]
  coordinates {(1,4.6587)(2,5.3322)(3,5.8381)(4,6.1367)(5,6.2517)(6,6.2517)(16,6.7484)(21,7.1442)(26,7.4844)(31,7.7038)(36,7.8175)(41,8.1807)(46,8.6127)(51,8.8566)(56,8.9519)(61,9.1532)(66,9.4219)(71,9.6439)(76,9.7743)(81,9.9634)(86,10.096)(91,10.2658)(96,10.415)(101,10.6164)};
    \addlegendentry{\footnotesize{T-NN TL 1\%}};
    
    \addplot[color=blue, mark=o, very thick]   
    coordinates {(1,5.596)(2,6.4538)(3,6.503)(8,8.9666)(10,10.1367)(15,11.0782)(20,11.633)(25,11.8368)(30,12.2481)(35,12.4994)(41,13.0165)(45,12.8452)(50,13.5576)(55,13.5576)(60,13.2952)(65,13.5576)(70,13.2952)(75,13.5576)(80,13.5576)(85,13.5576)(90,13.5576)(95,13.5576)(100,13.5576)};
    \addlegendentry{\footnotesize{T-NN w/o TL}};
    
    \addplot[color=violet, mark=*, very thick]     
    coordinates {(1,4.4712)(2,4.4712)(3,4.4712)(5,4.4712)(10,4.4712)(15,4.4712)(20,4.4712)(25,4.4712)(30,4.4712)(35,4.4712)(41,4.4712)(45,4.4712)(50,4.4712)(55,4.4712)(60,4.4712)(65,4.4712)(70,4.4712)(75,4.4712)(80,4.4712)(85,4.4712)(90,4.4712)(95,4.4712)(100,4.4712)};
    \addlegendentry{\footnotesize{S-NN}};

    \addplot[color=orange, mark=triangle, very thick]     
    coordinates {(1,6.5197)(2,6.5197)(3,6.5197)(8,6.5197)(10,6.5197)(15,6.5197)(20,6.5197)(25,6.5197)(30,6.5197)(35,6.5197)(41,6.5197)(45,6.5197)(50,6.5197)(55,6.5197)(60,6.5197)(65,6.5197)(70,6.5197)(75,6.5197)(80,6.5197)(85,6.5197)(90,6.5197)(95,6.5197)(100,6.5197)};
    \addlegendentry{\footnotesize{w/o NN}};
  \end{axis}
    \end{tikzpicture}
    \caption{ }
\end{subfigure}
\caption{Transfering the learning between the launch powers. Case I: from 8~dBm to (a) 7~dBm, (b) 6~dBm, (c) 5~dBm, using 18$\times$50~km SSMF fiber link and DP-16-QAM 34.4~GBd. Case II: from 5~dBm to (d) 4~dBm, (e) 3~dBm, (f) 2~dBm, using 9$\times$50~km TWC fiber link and DP-16-QAM 34.4~GBd. }
\label{fig:transfer_power}
\end{figure*}
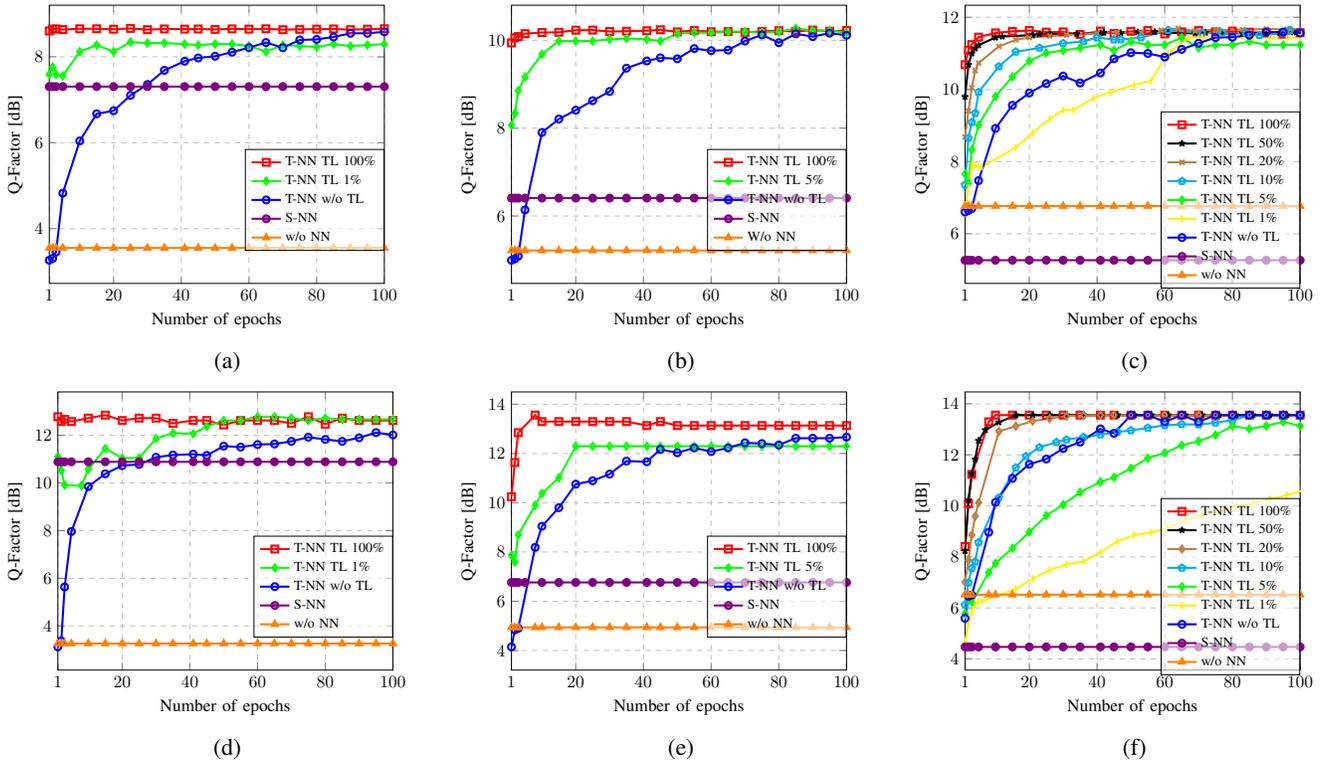

In general, for the CNN+biLSTM NN considered in this paper, we incorporate the mean square error (MSE) loss estimator and the classical Adam algorithm for the stochastic optimization step with the default learning rate set equal to 0.001~\cite{gulli2017deep}. The training was carried out for up to 200 epochs with a batch size of 1000, which has proven to be high enough to show the convergence for our transmission scenarios. Additionally, the total dataset used was composed of $2^{18}$ symbols for the training dataset and of $2^{18}$ independently generated symbols for the testing phase. The training dataset was shuffled at the beginning of every epoch to avoid overfitting caused by learning the connections between the neighboring training pairs~\cite{eriksson2017applying}. All datasets were generated using the Mersenne twister generator~\cite{matsumoto1998mersenne} with different random seeds, which guarantees a cross-correlation below 0.004 between the training and testing datasets, meaning that the symbols are virtually independent.

Finally, since the goal of this paper is to demonstrate the efficiency of the TL technique, we use the same best-performing CNN+biLSTM architecture with 244 filters, kernel size 10, and 226 hidden units in the LSTM cell, as in Ref.~\cite{freire2021performance}. Also, the number of taps used was $N = 40$: this is the maximal memory size estimation for all scenario changes that we will address. We notice that the memory effect is important because, even though we compensate the chromatic dispersion electronically, we still have to mitigate also the impact of the coupling between the nonlinearity and the chromatic dispersion along with optical fiber transmission. To unroll this coupling efficiently, we need the information from the neighboring symbols. Note that this is true in other perturbation techniques \cite{zhang2019field}, where the triplets are used to enhance the signal after the chromatic dispersion compensation. This memory guarantees that we do not artificially degrade the NN performance for all the cases considered. Also, we would like to highlight that the objective of the approach proposed in this paper was to produce a NN complex enough to deal with the different levels of nonlinearity but without needing to increase its number of hyperparameters (e.g. number of layers, neurons, filters, etc.). As it was explained in Ref.~\cite{freire2021performance}, increasing the NN parameters' number above the NN's capacity would cause overfitting, therefore limiting the achievable performance improvement. Clearly, in the ideal scenario, the neural network topology/hyperparameters should be optimized to the new configuration when changing the transmission setup. However, we decided to investigate in this work a more practical approach where the equalizer is kept unchanged although this approach requires higher complexity to fully cope with a strong nonlinear scenario.

\subsection{Transfer learning for different scenarios of launch power}\label{subSec:power}

We begin our analysis by transferring the learned parameters from a system that has been trained with a specific launch power to a system that operates at different power levels. Both the SSMF and TWC fiber types are considered. Fig.~\ref{fig:transfer_power} presents the results of TL between different powers for two cases, considering the SSMF and TWC separately. The first case (Case I) compares the system performance in terms of Q-factor when the source dataset consists of a 16-QAM signal with a launch power of 8~dBm to the three target datasets with different optical power: (a) 7~dBm, (b) 6~dBm, and (c) 5~dBm. The systems keep the same DP-16-QAM at 34.4~GBd, and the same transmission parameters (18$\times$50~km SSMF fiber link). Some important conclusions can be drawn from this figure. First, as expected, when moving from the source power (8~dBm) to the target power (5~dBm), the S-NN's output degraded showing even worse performance than the reference case (when only the linear equalization was applied). Since the conditional probability defined by the models did not generalize for different powers, this observation reveals that the NNs, by default, are not flexible enough to be used when the launch power changes. In practice, such changes in launch power may be necessitated by the addition of neighboring channels to the network, which will result in a reduction in the optimum launch power owing to XPM effects. However, we can see that the TL usage allows the NN to quickly reconfigure, and the latter then provides an efficient output in the new scenario. By retraining only the convolutional layers, we required 1 epoch, 4 epochs, and 10 epochs for the 7~dBm, 6~dBm, and 5~dBm scenarios, respectively, to achieve the best Q-factor. This translates into an approximate reduction in the number of epochs to 99\%, 95\%, and 88\%, respectively, compared to achieving the same Q-factor when trained from scratch. Another advantage of the TL is that the size of the required training dataset can be reduced without compromising the equalizer's efficiency. Case I shows that, depending on the difference between the launch power considered in the source and target links, we can save up to 99 \% of the amount of training data. Note that we carried out the analysis when training with different dataset sizes (1/ 5/ 10/ 20/ 50/ 100\%), as shown in Fig.~\ref{fig:transfer_power} (c) and (f). However, for clarity, in the rest of this paper, we only report the TL for the cases where we saved the most epochs using 100\% of the dataset, and the case in which we trained with the least amount of data while achieving the same or better Q-factor as in the case without the TL.

Case II illustrates how general our findings are. We checked a source dataset with DP-16-QAM 34.4~GBd considering a launch power of 5~dBm in a 9$\times$50~km TWC link and transferred it to the target sets of (d) 4~dB, (e) 3~dBm, and (f) 2~dBm launch powers, with the same setup. We can see that the use of TL was helpful in this case as well, and the results are given in Fig~\ref{fig:transfer_power}. When switching to (d) 4~dBm, (e) 3~dBm, and (f) 2~dBm, the number of epochs necessary to achieve the maximum Q-factor decreased approximately by 99\%, 90\%, and 80\%, respectively. Also, we can see that the re-training process required fewer data: 99\%, 95\%, and 95\%, respectively. 

At this stage, it is pertinent to address three questions: i) what happens if the launch power is reduced further; ii) can the knowledge be transferred from lower to higher launch powers; and iii) would the TL still work in the presence of a transceiver noise. To answer the first question, we tested the TL for a wide range of launch power levels (from 8 to -8~dBm) and found that it performs quite well as long as the NN-based equalizer still works, i.e., for the powers where it produces a non-zero improvement in symbol recovery. The TL works well in the nonlinear fiber transmission regime because the NN reverses the Kerr nonlinearity and uncompensated dispersion-related effects. However, the equalizer's effectiveness degrades in the linear regime. This is because the NN equalizer itself cannot recover impairments in the linear regime of our simulated data, since they come from a stochastic noise-related effect emerging from the amplifiers. Note that in the simulations we considered the ideal components, e.g. DAC/ADC, etc. Nonetheless, despite the fact that we use such high power, 7/ 6/ 5~dBm, the TL works efficiently in the whole range of powers down to the optimal launch power.

Question ii) is relevant because it stresses the need of comprehending the underlying physical effects. After analyzing the effectiveness of transferring the knowledge from smaller to higher launch powers and vice-versa, we found that the \textit{TL is more effective when the training is carried out at higher launch powers and the TL occurs from higher to lower launch powers}. The results regarding the TL direction effects are summarized in Table~\ref{Table_TL_power}. The explanation for the TL direction dependence is that the NN equalizer reverses the nonlinear effects, and the latter intensify with the growth of launch power. The source domain NN trained with higher launch power possesses a better ``knowledge'' of nonlinearity, the NN  learns the nonlinear channel function well. Therefore, the high power regime has more useful information about the channel function, which can be readily adapted for target systems with lower launch powers, compared to the situation when we employ the TL in the opposite direction. To illustrate this phenomenon, consider the case where a polynomial Volterra equalizer is used. It is evident that if we train it for a higher nonlinearity scenario, we will need more coefficients to get a satisfactory result than if we train it for a lower nonlinearity \cite{sena2021bayesian}. The TL follows the same logic whereby training the source NN with the worst-case scenario, it is easier for NN architectures to discard some of the source NN's elements that are not meaningful in the lower nonlinear regime than it is to learn new ones from scratch.

\begin{table}[htbp] 
  \centering
  \caption{Dependence of the TL performance on the transfer direction, for the case where we change only the launch powers from the source to target datasets.}
\begin{tabular}{|c|c|c|c|}
\hline
Fiber                 & Scenario         & Max Q-factor & Epochs required \\ \hline \hline
\multirow{3}{*}{SSMF} & TL 8 dBm $\rightarrow$ 5dBm  &        11.56      &    \textless 10    \\ \cline{2-4} 
                      & TL 2 dBm $\rightarrow$ 5dBm  &       11.56       &   \textgreater 100     \\ \cline{2-4} 
                      & w/o TL           &          11.56    &     \textgreater 80   \\ \hline\hline
\multirow{3}{*}{TWC}  & TL 5 dBm $\rightarrow$ 2dB   &        13.56      &   \textless 10    \\ \cline{2-4} 
                      & TL -1 dBm $\rightarrow$ 2dBm &      13.56        &  \textgreater 100      \\ \cline{2-4} 
                      & w/o TL           &          13.56    &    \textgreater 45    \\ \hline
\end{tabular}
\label{Table_TL_power}
\end{table}

 \begin{figure*}[ht!]
 \centering
    \begin{subfigure}[b]{.45\textwidth}
    \centering
  \begin{tikzpicture}[scale=0.7]
    \begin{axis} [
        xlabel={Number of epochs},
        ylabel={Q-Factor [dB]},
        grid=both,          xtick = {1,20,40,60,80,100},   
    	xmin=1, xmax=100,
        legend style={legend style={ at={(1,0.3)},anchor= east}, legend cell align=left,fill=white, fill opacity=0.6, draw opacity=1,text opacity=1},
    	grid style={dashed}]
        ]
      \addplot[color=red, mark=square, very thick]
    coordinates {(1,8.0359)(2,8.2061)(3,8.2653)(5,8.3107)(10,8.3482)(15,8.3485)(20,8.3651)(25,8.3682)(30,8.3548)(35,8.3577)(41,8.3643)(45,8.3468)(50,8.3692)(55,8.3692)(60,8.3692)(65,8.3692)(70,8.3692)(75,8.3692)(80,8.3692)(85,8.3692)(90,8.3692)(95,8.3692)(100,8.3692)};
    \addlegendentry{\footnotesize{T-NN TL 100\%}};

    \addplot[color=green, mark=diamond, very thick]
  coordinates {(1,5.6836)(2,6.1906)(3,6.447)(5,6.3937)(10,6.5106)(15,6.9882)(20,7.0689)(25,7.2273)(30,7.3718)(35,7.588)(41,7.6982)(45,7.7814)(50,7.8332)(55,7.8687)(60,7.9233)(65,7.9049)(70,7.999)(75,7.9798)(80,8.0185)(85,8.0784)(90,8.0784)(95,8.162)(100,8.1407)};
    \addlegendentry{\footnotesize{T-NN TL 1\%}};
    
    \addplot[color=blue, mark=o, very thick]   
    coordinates {(1,5.2766)(2,5.3664)(3,5.3871)(5,5.8256)(10,7.1254)(15,7.3548)(20,7.4473)(25,7.5753)(30,7.6119)(35,7.7355)(41,7.8755)(45,7.9064)(50,7.9436)(55,7.9231)(60,7.9848)(65,7.9936)(70,8.0293)(75,7.8285)(80,7.9664)(85,7.9589)(90,7.9668)(95,7.9053)(100,7.7452)};
    \addlegendentry{\footnotesize{T-NN w/o TL}};
    
    \addplot[color=violet, mark=*, very thick]     
    coordinates {(1,3.2967)(2,3.2967)(3,3.2967)(5,3.2967)(10,3.2967)(15,3.2967)(20,3.2967)(25,3.2967)(30,3.2967)(35,3.2967)(41,3.2967)(45,3.2967)(50,3.2967)(55,3.2967)(60,3.2967)(65,3.2967)(70,3.2967)(75,3.2967)(80,3.2967)(85,3.2967)(90,3.2967)(95,3.2967)(100,3.2967)};
    \addlegendentry{\footnotesize{S-NN}};

    \addplot[color=orange, mark=triangle, very thick]     
    coordinates {(1,5.5488)(2,5.5488)(3,5.5488)(5,5.5488)(10,5.5488)(15,5.5488)(20,5.5488)(25,5.5488)(30,5.5488)(35,5.5488)(41,5.5488)(45,5.5488)(50,5.5488)(55,5.5488)(60,5.5488)(65,5.5488)(70,5.5488)(75,5.5488)(80,5.5488)(85,5.5488)(90,5.5488)(95,5.5488)(100,5.5488)};
    \addlegendentry{\footnotesize{w/o NN}};
  \end{axis}
    \end{tikzpicture}
    \caption{ }
    \end{subfigure}
	\begin{subfigure}[b]{.45\textwidth}
    \centering
  \begin{tikzpicture}[scale=0.7]
    \begin{axis} [
        xlabel={Number of epochs},
        ylabel={Q-Factor [dB]},
        grid=both,          xtick = {1,20,40,60,80,100},   
    	xmin=1, xmax=100,
        legend style={legend style={ at={(1,0.3)},anchor= east}, legend cell align=left,fill=white, fill opacity=0.6, draw opacity=1,text opacity=1},
    	grid style={dashed}]
        ]
      \addplot[color=red, mark=square, very thick]
    coordinates {(1,8.3492)(2,9.2627)(3,9.523)(5,9.6952)(10,9.7612)(15,9.7916)(20,9.8294)(25,9.8204)(30,9.8059)(35,9.8245)(41,9.8019)(45,9.8155)(50,9.8035)(55,9.8212)(60,9.8269)(65,9.822)(70,9.8051)(75,9.8196)(80,9.8123)(85,9.8245)(90,9.8115)(95,9.8277)(100,9.818)};
    \addlegendentry{\footnotesize{T-NN TL 100\%}};

   \addplot[color=green, mark=diamond, very thick]
  coordinates {(1,5.2647)(2,5.4276)(3,5.7101)(5,6.1909)(10,7.2659)(15,8.0909)(20,8.5515)(25,8.9655)(30,9.2754)(35,9.4444)(41,9.5905)(45,9.7147)(50,9.5905)(55,9.6577)(60,9.7751)(65,9.7002)(70,9.6577)(75,9.6577)(80,9.7597)(85,9.7597)(90,9.7908)(95,9.7751)(100,9.7295)};

    \addlegendentry{\footnotesize{T-NN TL 5\%}};
    
    \addplot[color=blue, mark=o, very thick]   
    coordinates {(1,5.0974)(2,5.2214)(3,5.7083)(5,7.7012)(10,8.5822)(15,9.0279)(20,9.2291)(25,9.3343)(30,9.3759)(35,9.4174)(41,9.4185)(45,9.5353)(50,9.5597)(55,9.4675)(60,9.5132)(65,9.509)(70,9.5648)(75,9.5609)(80,9.3834)(85,9.509)(90,9.509)(95,9.509)(100,9.509)};
    \addlegendentry{\footnotesize{T-NN w/o TL}};
    
    \addplot[color=violet, mark=*, very thick]     
    coordinates {(1,3.6618)(2,3.6618)(3,3.6618)(5,3.6618)(10,3.6618)(15,3.6618)(20,3.6618)(25,3.6618)(30,3.6618)(35,3.6618)(41,3.6618)(45,3.6618)(50,3.6618)(55,3.6618)(60,3.6618)(65,3.6618)(70,3.6618)(75,3.6618)(80,3.6618)(85,3.6618)(90,3.6618)(95,3.6618)(100,3.6618)};
    \addlegendentry{\footnotesize{S-NN}};
    \addplot[color=orange, mark=triangle, very thick]  
    coordinates {(1,5.1861)(2,5.1861)(3,5.1861)(5,5.1861)(10,5.1861)(15,5.1861)(20,5.1861)(25,5.1861)(30,5.1861)(35,5.1861)(41,5.1861)(45,5.1861)(50,5.1861)(55,5.1861)(60,5.1861)(65,5.1861)(70,5.1861)(75,5.1861)(80,5.1861)(85,5.1861)(90,5.1861)(95,5.1861)(100,5.1861)};
    \addlegendentry{\footnotesize{W/o NN}};
  \end{axis}
    \end{tikzpicture}
    \caption{ }
    \end{subfigure}

\caption{\label{NoisePower}Launch power transfer learning from a dataset without transmitter noise to the dataset with transmitter noise. (a) SSMF (from 8~dBm to 5~dBm); (b) TWC (from 5~dBm to 2~dBm).}
\end{figure*}
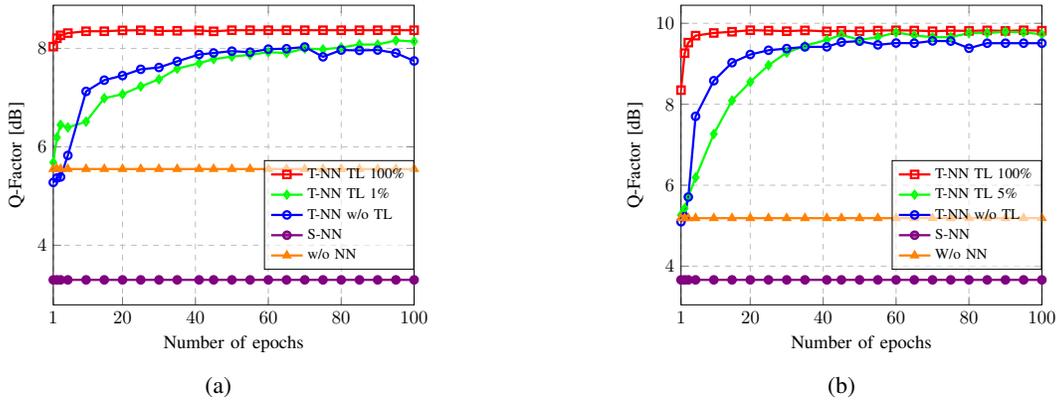

Finally, we added Fig.~\ref{NoisePower} to answer the third question of whether the TL would still work with the addition of component-generated noise. To generate the data with noise, we assumed realistic transceivers affected by electrical noise with back-to-back SNR given by:
\begin{equation}
    \text{SNR[dB]} = -0.175 R + 30,
\end{equation}
where $R$ is the symbol rate\footnote{This equation was derived by the authors of Ref.~\cite{8386004} for a reference system and distributed within TRANSNET Project members.}. This equation is an approximate fit to the experimentally measured values described in~\cite{8386004} and the noise, modeled as an additive white Gaussian, is contributed equally by the transmitter and receiver. Fig.~\ref{NoisePower} (a) refers to the case of 18$\times$50~km SSMF transferring from the source 8~dBm (without transceiver noise) to the target 5~dBm (with transceiver noise); Fig.~\ref{NoisePower} (b) refers to the 9$\times$50~km TWC transferring from the source 5~dBm (without transceiver noise) to the target 2~dBm (with transceiver noise). The analysis of the results given in Fig.~\ref{NoisePower} reveals that the system performance was slightly impacted by the increased noise level, but the TL continued to work with the same effectiveness. In Fig.~\ref{NoisePower} (a), the reduction of $ 90$\% in epochs and 99\% in the training dataset was observed. By the same token, Fig.~\ref{NoisePower} (b) shows the decrease of $ 90$\% in epochs and 95\% of the training dataset. Note that the case depicted in Fig.~\ref{NoisePower} was selected to highlight that the transmitter noise in the target domain does not harm the TL performance. We have also tested adding noise to both the source and target domains with the TL still showing quite good performance in this case as well.

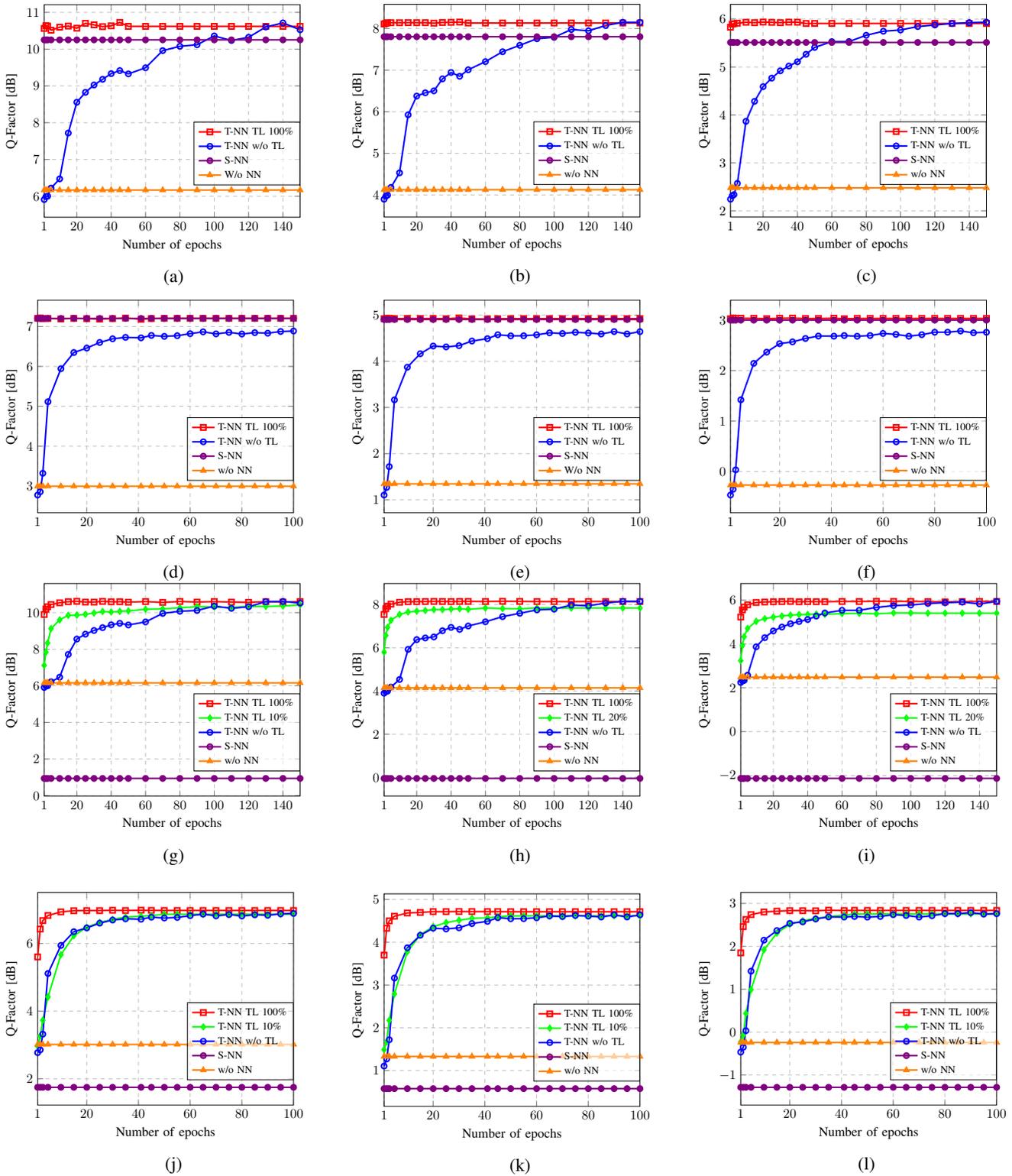
\begin{figure*}[ht!]
    \centering 
\begin{subfigure}{0.33\textwidth}
  \begin{tikzpicture}[scale=0.65]
    \begin{axis} [
        xlabel={Number of epochs},
        ylabel={Q-Factor [dB]},
        grid=both,          xtick = {1,20,40,60,80,100,120,140},   
    	xmin=1, xmax=150,
        legend style={legend style={ at={(1,0.3)},anchor= east}, legend cell align=left,fill=white, fill opacity=0.6, draw opacity=1,text opacity=1},
    	grid style={dashed}]
        ]
      \addplot[color=red, mark=square, very thick]
    coordinates {(1,10.5628)(2,10.6369)(3,10.6263)(5,10.5129)(10,10.5953)(15,10.6263)(20,10.569)(25,10.7021)(30,10.6599)(35,10.6106)(40,10.6329)(45,10.7254)(50,10.6184)(60,10.6184)(70,10.6184)(80,10.6184)(90,10.6184)(100,10.6184)(110,10.6184)(120,10.6184)(130,10.6184)(140,10.6184)(150,10.6184)};
    \addlegendentry{\footnotesize{T-NN TL 100\%}};
    
    \addplot[color=blue, mark=o, very thick]   
    coordinates {(1,5.9097)(2,5.9911)(3,6.0024)(5,6.2223)(10,6.472)(15,7.718)(20,8.5557)(25,8.825)(30,9.0247)(35,9.1796)(40,9.3338)(45,9.4161)(50,9.3272)(60,9.4929)(70,9.9595)(80,10.077)(90,10.1193)(100,10.3583)(110,10.235)(120,10.3222)(130,10.601)(140,10.7121)(150,10.5251)};
    \addlegendentry{\footnotesize{T-NN w/o TL}};
    
    \addplot[color=violet, mark=*, very thick]     
    coordinates {(1,10.2518)(2,10.2518)(3,10.2518)(5,10.2518)(10,10.2518)(15,10.2518)(20,10.2518)(25,10.2518)(30,10.2518)(35,10.2518)(40,10.2518)(45,10.2518)(50,10.2518)(60,10.2518)(70,10.2518)(80,10.2518)(90,10.2518)(100,10.2518)(110,10.2518)(120,10.2518)(130,10.2518)(140,10.2518)(150,10.2518)};
    \addlegendentry{\footnotesize{S-NN}};

    \addplot[color=orange, mark=triangle, very thick]  coordinates {(1,6.1692)(2,6.1692)(3,6.1692)(5,6.1692)(10,6.1692)(15,6.1692)(20,6.1692)(25,6.1692)(30,6.1692)(35,6.1692)(40,6.1692)(45,6.1692)(50,6.1692)(60,6.1692)(70,6.1692)(80,6.1692)(90,6.1692)(100,6.1692)(110,6.1692)(120,6.1692)(130,6.1692)(140,6.1692)(150,6.1692)};

    \addlegendentry{\footnotesize{W/o NN}};
    
  \end{axis}
    \end{tikzpicture}
  \caption{ }
\end{subfigure}\hfil 
\begin{subfigure}{0.33\textwidth}
  \begin{tikzpicture}[scale=0.65]
    \begin{axis} [
        xlabel={Number of epochs},
        ylabel={Q-Factor [dB]},
        grid=both,          xtick = {1,20,40,60,80,100,120,140},   
    	xmin=1, xmax=150,
        legend style={legend style={ at={(1,0.3)},anchor= east}, legend cell align=left,fill=white, fill opacity=0.6, draw opacity=1,text opacity=1},
    	grid style={dashed}]
        ]
      \addplot[color=red, mark=square, very thick]
    coordinates {(1,8.1096)(2,8.126)(3,8.1389)(5,8.1392)(10,8.139)(15,8.1411)(20,8.1398)(25,8.1402)(30,8.1293)(35,8.1455)(40,8.1492)(45,8.1553)(50,8.1326)(60,8.1326)(70,8.1326)(80,8.1326)(90,8.1326)(100,8.1326)(110,8.1326)(120,8.1326)(130,8.1326)(140,8.1326)(150,8.1326)};
    \addlegendentry{\footnotesize{T-NN TL 100\%}};

    \addplot[color=blue, mark=o, very thick]   
    coordinates {(1,3.897)(2,3.9688)(3,3.9938)(5,4.1809)(10,4.5306)(15,5.926)(20,6.3766)(25,6.4551)(30,6.5021)(35,6.79)(40,6.9436)(45,6.8524)(50,7.0088)(60,7.2046)(70,7.4427)(80,7.5971)(90,7.755)(100,7.7891)(110,7.9772)(120,7.946)(130,8.0728)(140,8.1494)(150,8.148)};
    \addlegendentry{\footnotesize{T-NN w/o TL}};
    
    \addplot[color=violet, mark=*, very thick]     
    coordinates {(1,7.8031)(2,7.8031)(3,7.8031)(5,7.8031)(10,7.8031)(15,7.8031)(20,7.8031)(25,7.8031)(30,7.8031)(35,7.8031)(40,7.8031)(45,7.8031)(50,7.8031)(60,7.8031)(70,7.8031)(80,7.8031)(90,7.8031)(100,7.8031)(110,7.8031)(120,7.8031)(130,7.8031)(140,7.8031)(150,7.8031)};
    \addlegendentry{\footnotesize{S-NN}};

    \addplot[color=orange, mark=triangle, very thick]     
    coordinates {(1,4.1256)(2,4.1256)(3,4.1256)(5,4.1256)(10,4.1256)(15,4.1256)(20,4.1256)(25,4.1256)(30,4.1256)(35,4.1256)(40,4.1256)(45,4.1256)(50,4.1256)(60,4.1256)(70,4.1256)(80,4.1256)(90,4.1256)(100,4.1256)(110,4.1256)(120,4.1256)(130,4.1256)(140,4.1256)(150,4.1256)};
    \addlegendentry{\footnotesize{w/o NN}};
  \end{axis}
    \end{tikzpicture}
  \caption{ }
\end{subfigure}\hfil 
\begin{subfigure}{0.33\textwidth}
  \begin{tikzpicture}[scale=0.65]
    \begin{axis} [
        xlabel={Number of epochs},
        ylabel={Q-Factor [dB]},
        grid=both,          xtick = {1,20,40,60,80,100,120,140},   
    	xmin=1, xmax=150,
        legend style={legend style={ at={(1,0.3)},anchor= east}, legend cell align=left,fill=white, fill opacity=0.6, draw opacity=1,text opacity=1},
    	grid style={dashed}]
        ]
      \addplot[color=red, mark=square, very thick]
    coordinates {(1,5.83)(2,5.8954)(3,5.9038)(5,5.9146)(10,5.9336)(15,5.9247)(20,5.9378)(25,5.9321)(30,5.9249)(35,5.936)(40,5.935)(45,5.9085)(50,5.9093)(60,5.9093)(70,5.9093)(80,5.9093)(90,5.9093)(100,5.9093)(110,5.9093)(120,5.9093)(130,5.9093)(140,5.9093)(150,5.9093)};
    \addlegendentry{\footnotesize{T-NN TL 100\%}};
    
    \addplot[color=blue, mark=o, very thick]   
    coordinates {(1,2.242)(2,2.3129)(3,2.3373)(5,2.5716)(10,3.8669)(15,4.2836)(20,4.5908)(25,4.7683)(30,4.9204)(35,5.0201)(40,5.1104)(45,5.2662)(50,5.4095)(60,5.5298)(70,5.5317)(80,5.6624)(90,5.7468)(100,5.7722)(110,5.8441)(120,5.8764)(130,5.9074)(140,5.9252)(150,5.9348)};
    \addlegendentry{\footnotesize{T-NN w/o TL}};
    
    \addplot[color=violet, mark=*, very thick]     
    coordinates {(1,5.5131)(2,5.5131)(3,5.5131)(5,5.5131)(10,5.5131)(15,5.5131)(20,5.5131)(25,5.5131)(30,5.5131)(35,5.5131)(40,5.5131)(45,5.5131)(50,5.5131)(60,5.5131)(70,5.5131)(80,5.5131)(90,5.5131)(100,5.5131)(110,5.5131)(120,5.5131)(130,5.5131)(140,5.5131)(150,5.5131)};
    \addlegendentry{\footnotesize{S-NN}};

    \addplot[color=orange, mark=triangle, very thick]  coordinates {(1,2.4791)(2,2.4791)(3,2.4791)(5,2.4791)(10,2.4791)(15,2.4791)(20,2.4791)(25,2.4791)(30,2.4791)(35,2.4791)(40,2.4791)(45,2.4791)(50,2.4791)(60,2.4791)(70,2.4791)(80,2.4791)(90,2.4791)(100,2.4791)(110,2.4791)(120,2.4791)(130,2.4791)(140,2.4791)(150,2.4791)};
    \addlegendentry{\footnotesize{w/o NN}};
  \end{axis}
    \end{tikzpicture}
  \caption{ }
\end{subfigure}

\medskip
\begin{subfigure}{0.33\textwidth}
  \begin{tikzpicture}[scale=0.65]
    \begin{axis} [
        xlabel={Number of epochs},
        ylabel={Q-Factor [dB]},
        grid=both,          xtick = {1,20,40,60,80,100},   
    	xmin=1, xmax=100,
        legend style={legend style={ at={(1,0.3)},anchor= east}, legend cell align=left,fill=white, fill opacity=0.6, draw opacity=1,text opacity=1},
    	grid style={dashed}]
        ]
      \addplot[color=red, mark=square, very thick]
    coordinates {(1,7.2056)(2,7.2157)(3,7.1985)(5,7.2026)(10,7.1908)(15,7.2055)(20,7.1975)(25,7.1857)(30,7.1994)(35,7.2131)(41,7.1814)(45,7.201)(50,7.2061)(55,7.2061)(60,7.2061)(65,7.2061)(70,7.2061)(75,7.2061)(80,7.2061)(85,7.2061)(90,7.2061)(95,7.2061)(100,7.2061)};
    \addlegendentry{\footnotesize{T-NN TL 100\%}};

    \addplot[color=blue, mark=o, very thick]   
    coordinates {(1,2.774)(2,2.8562)(3,3.3204)(5,5.1136)(10,5.9427)(15,6.3477)(20,6.4597)(25,6.6003)(30,6.6927)(35,6.7282)(41,6.7164)(45,6.7769)(50,6.7532)(55,6.77)(60,6.8208)(65,6.8667)(70,6.8159)(75,6.8537)(80,6.8124)(85,6.8469)(90,6.8327)(95,6.8753)(100,6.8876)};
    \addlegendentry{\footnotesize{T-NN w/o TL}};
    
    \addplot[color=violet, mark=*, very thick]     
   coordinates {(1,7.2036)(2,7.2036)(3,7.2036)(5,7.2036)(10,7.2036)(15,7.2036)(20,7.2036)(25,7.2036)(30,7.2036)(35,7.2036)(41,7.2036)(45,7.2036)(50,7.2036)(55,7.2036)(60,7.2036)(65,7.2036)(70,7.2036)(75,7.2036)(80,7.2036)(85,7.2036)(90,7.2036)(95,7.2036)(100,7.2036)};
    \addlegendentry{\footnotesize{S-NN}};

    \addplot[color=orange, mark=triangle, very thick]     
    coordinates {(1,2.9961)(2,2.9961)(3,2.9961)(5,2.9961)(10,2.9961)(15,2.9961)(20,2.9961)(25,2.9961)(30,2.9961)(35,2.9961)(41,2.9961)(45,2.9961)(50,2.9961)(55,2.9961)(60,2.9961)(65,2.9961)(70,2.9961)(75,2.9961)(80,2.9961)(85,2.9961)(90,2.9961)(95,2.9961)(100,2.9961)};
    
    \addlegendentry{\footnotesize{w/o NN}};
  \end{axis}
    \end{tikzpicture}
  \caption{ }
\end{subfigure}\hfil 
\begin{subfigure}{0.33\textwidth}
  \begin{tikzpicture}[scale=0.65]
    \begin{axis} [
        xlabel={Number of epochs},
        ylabel={Q-Factor [dB]},
        grid=both,          xtick = {1,20,40,60,80,100},   
    	xmin=1, xmax=100,
        legend style={legend style={ at={(1,0.3)},anchor= east}, legend cell align=left,fill=white, fill opacity=0.6, draw opacity=1,text opacity=1},
    	grid style={dashed}]
        ]
      \addplot[color=red, mark=square, very thick]
    coordinates {(1,4.9219)(2,4.9177)(3,4.9293)(5,4.9227)(10,4.9273)(15,4.9254)(20,4.9261)(25,4.9205)(30,4.9376)(35,4.9177)(41,4.9152)(45,4.9211)(50,4.923)(55,4.923)(60,4.923)(65,4.923)(70,4.923)(75,4.923)(80,4.923)(85,4.923)(90,4.923)(95,4.923)(100,4.923)};
    \addlegendentry{\footnotesize{T-NN TL 100\%}};
    
    \addplot[color=blue, mark=o, very thick]   
    coordinates {(1,1.104)(2,1.269)(3,1.7198)(5,3.1636)(10,3.8704)(15,4.1631)(20,4.3295)(25,4.3108)(30,4.3373)(35,4.4376)(41,4.4888)(45,4.5741)(50,4.5496)(55,4.5515)(60,4.5721)(65,4.6151)(70,4.6004)(75,4.626)(80,4.6122)(85,4.5904)(90,4.6417)(95,4.5912)(100,4.6405)};
    \addlegendentry{\footnotesize{T-NN w/o TL}};
    
    \addplot[color=violet, mark=*, very thick]     
    coordinates {(1,4.9036)(2,4.9036)(3,4.9036)(5,4.9036)(10,4.9036)(15,4.9036)(20,4.9036)(25,4.9036)(30,4.9036)(35,4.9036)(41,4.9036)(45,4.9036)(50,4.9036)(55,4.9036)(60,4.9036)(65,4.9036)(70,4.9036)(75,4.9036)(80,4.9036)(85,4.9036)(90,4.9036)(95,4.9036)(100,4.9036)};
    \addlegendentry{\footnotesize{S-NN}};

    \addplot[color=orange, mark=triangle, very thick]
    coordinates {(1,1.3465)(2,1.3465)(3,1.3465)(5,1.3465)(10,1.3465)(15,1.3465)(20,1.3465)(25,1.3465)(30,1.3465)(35,1.3465)(41,1.3465)(45,1.3465)(50,1.3465)(55,1.3465)(60,1.3465)(65,1.3465)(70,1.3465)(75,1.3465)(80,1.3465)(85,1.3465)(90,1.3465)(95,1.3465)(100,1.3465)};
    \addlegendentry{\footnotesize{W/o NN}};
  \end{axis}
    \end{tikzpicture}
  \caption{ }
\end{subfigure}\hfil 
\begin{subfigure}{0.33\textwidth}
  \begin{tikzpicture}[scale=0.65]
    \begin{axis} [
        xlabel={Number of epochs},
        ylabel={Q-Factor [dB]},
        grid=both,          xtick = {1,20,40,60,80,100},   
    	xmin=1, xmax=100,
        legend style={legend style={ at={(1,0.3)},anchor= east}, legend cell align=left,fill=white, fill opacity=0.6, draw opacity=1,text opacity=1},
    	grid style={dashed}]
        ]
      \addplot[color=red, mark=square, very thick]
    coordinates {(1,3.0279)(2,3.0443)(3,3.0354)(5,3.0385)(10,3.037)(15,3.0301)(20,3.0379)(25,3.0416)(30,3.0334)(35,3.0341)(41,3.0392)(45,3.0284)(50,3.0354)(55,3.0354)(60,3.0354)(65,3.0354)(70,3.0354)(75,3.0354)(80,3.0354)(85,3.0354)(90,3.0354)(95,3.0354)(100,3.0354)};
    \addlegendentry{\footnotesize{T-NN TL 100\%}};
    \addplot[color=blue, mark=o, very thick]   
    coordinates {(1,-0.46653)(2,-0.35225)(3,0.032936)(5,1.4213)(10,2.1435)(15,2.3663)(20,2.5351)(25,2.5695)(30,2.6371)(35,2.6855)(41,2.6829)(45,2.6947)(50,2.6812)(55,2.6954)(60,2.7369)(65,2.718)(70,2.6822)(75,2.7122)(80,2.7607)(85,2.7617)(90,2.7853)(95,2.7503)(100,2.7604)};
    \addlegendentry{\footnotesize{T-NN w/o TL}};
    
    \addplot[color=violet, mark=*, very thick]     
    coordinates {(1,3.0006)(2,3.0006)(3,3.0006)(5,3.0006)(10,3.0006)(15,3.0006)(20,3.0006)(25,3.0006)(30,3.0006)(35,3.0006)(41,3.0006)(45,3.0006)(50,3.0006)(55,3.0006)(60,3.0006)(65,3.0006)(70,3.0006)(75,3.0006)(80,3.0006)(85,3.0006)(90,3.0006)(95,3.0006)(100,3.0006)};
    \addlegendentry{\footnotesize{S-NN}};

    \addplot[color=orange, mark=triangle, very thick]  
    coordinates {(1,-0.26928)(2,-0.26928)(3,-0.26928)(5,-0.26928)(10,-0.26928)(15,-0.26928)(20,-0.26928)(25,-0.26928)(30,-0.26928)(35,-0.26928)(41,-0.26928)(45,-0.26928)(50,-0.26928)(55,-0.26928)(60,-0.26928)(65,-0.26928)(70,-0.26928)(75,-0.26928)(80,-0.26928)(85,-0.26928)(90,-0.26928)(95,-0.26928)(100,-0.26928)};
    \addlegendentry{\footnotesize{w/o NN}};
  \end{axis}
    \end{tikzpicture}
  \caption{ }
\end{subfigure}
\medskip
\begin{subfigure}{0.33\textwidth}
  \begin{tikzpicture}[scale=0.65]
    \begin{axis} [
        xlabel={Number of epochs},
        ylabel={Q-Factor [dB]},
        grid=both,          xtick = {1,20,40,60,80,100,120,140},   
    	xmin=1, xmax=150,
        legend style={legend style={ at={(1,0.3)},anchor= east}, legend cell align=left,fill=white, fill opacity=0.6, draw opacity=1,text opacity=1},
    	grid style={dashed}]
        ]
         \addplot[color=red, mark=square, very thick]
      coordinates {(1,9.8989)(2,10.2)(3,10.3228)(5,10.4435)(10,10.539)(15,10.5961)(20,10.62)(25,10.5791)(30,10.5781)(35,10.6167)(40,10.5929)(45,10.595)(50,10.5728)(60,10.6047)(70,10.5573)(80,10.6113)(90,10.5791)(100,10.6036)(110,10.5781)(120,10.5624)(130,10.5972)(140,10.5812)(150,10.6026)};
    \addlegendentry{\footnotesize{T-NN TL 100\%}};

    \addplot[color=green, mark=diamond, very thick]
    coordinates {(1,7.1166)(2,7.8376)(3,8.3431)(5,9.1295)(10,9.6036)(15,9.8526)(20,9.8661)(25,9.9078)(30,9.9814)(35,10.0611)(40,10.0365)(45,10.0694)(50,10.0949)(60,10.1852)(70,10.2044)(80,10.2645)(90,10.3289)(100,10.3069)(110,10.2854)(120,10.3401)(130,10.3401)(140,10.3629)(150,10.4227)};

    \addlegendentry{\footnotesize{T-NN TL 10\%}};
    
    \addplot[color=blue, mark=o, very thick]   
    coordinates {(1,5.9097)(2,5.9911)(3,6.0024)(5,6.2223)(10,6.472)(15,7.718)(20,8.5557)(25,8.825)(30,9.0247)(35,9.1796)(40,9.3338)(45,9.4161)(50,9.3272)(60,9.4929)(70,9.9595)(80,10.077)(90,10.1193)(100,10.3583)(110,10.235)(120,10.3222)(130,10.601)(140,10.6121)(150,10.5251)};
    \addlegendentry{\footnotesize{T-NN w/o TL}};
    
    \addplot[color=violet, mark=*, very thick]     
    coordinates {(1,0.94031)(2,0.94031)(3,0.94031)(5,0.94031)(10,0.94031)(15,0.94031)(20,0.94031)(25,0.94031)(30,0.94031)(35,0.94031)(40,0.94031)(45,0.94031)(50,0.94031)(60,0.94031)(70,0.94031)(80,0.94031)(90,0.94031)(100,0.94031)(110,0.94031)(120,0.94031)(130,0.94031)(140,0.94031)(150,0.94031)};
    \addlegendentry{\footnotesize{S-NN}};

    \addplot[color=orange, mark=triangle, very thick]  coordinates {(1,6.1604)(2,6.1604)(3,6.1604)(5,6.1604)(10,6.1604)(15,6.1604)(20,6.1604)(25,6.1604)(30,6.1604)(35,6.1604)(40,6.1604)(45,6.1604)(50,6.1604)(60,6.1604)(70,6.1604)(80,6.1604)(90,6.1604)(100,6.1604)(110,6.1604)(120,6.1604)(130,6.1604)(140,6.1604)(150,6.1604)};
    \addlegendentry{\footnotesize{w/o NN}};
  \end{axis}
    \end{tikzpicture}
  \caption{ }
\end{subfigure}\hfil 
\begin{subfigure}{0.33\textwidth}
  \begin{tikzpicture}[scale=0.65]
    \begin{axis} [
        xlabel={Number of epochs},
        ylabel={Q-Factor [dB]},
        grid=both,          xtick = {1,20,40,60,80,100,120,140},   
    	xmin=1, xmax=150,
        legend style={legend style={ at={(1,0.3)},anchor= east}, legend cell align=left,fill=white, fill opacity=0.6, draw opacity=1,text opacity=1},
    	grid style={dashed}]
        ]
      \addplot[color=red, mark=square, very thick]
    coordinates {(1,7.5433)(2,7.8118)(3,7.9246)(5,8.022)(10,8.1088)(15,8.1333)(20,8.1345)(25,8.1421)(30,8.1481)(35,8.1443)(40,8.1392)(45,8.146)(50,8.1476)(60,8.143)(70,8.1524)(80,8.1475)(90,8.1401)(100,8.1401)(110,8.1368)(120,8.1422)(130,8.1422)(140,8.1422)(150,8.1422)};
    \addlegendentry{\footnotesize{T-NN TL 100\%}};

    \addplot[color=green, mark=diamond, very thick]
  coordinates {(1,5.7972)(2,6.5718)(3,6.943)(5,7.2724)(10,7.5444)(15,7.6583)(20,7.6957)(25,7.7274)(30,7.7691)(35,7.7674)(40,7.7918)(45,7.8079)(50,7.7816)(60,7.8416)(70,7.8165)(80,7.8067)(90,7.8416)(100,7.8416)(110,7.8416)(120,7.8416)(130,7.8416)(140,7.8416)(150,7.8416)};
    \addlegendentry{\footnotesize{T-NN TL 20\%}};

    \addplot[color=blue, mark=o, very thick]   
   coordinates {(1,3.897)(2,3.9688)(3,3.9938)(5,4.1809)(10,4.5306)(15,5.926)(20,6.3766)(25,6.4551)(30,6.5021)(35,6.79)(40,6.9436)(45,6.8524)(50,7.0088)(60,7.2046)(70,7.4427)(80,7.5971)(90,7.755)(100,7.7891)(110,7.9772)(120,7.946)(130,8.0728)(140,8.1494)(150,8.1494)};
    \addlegendentry{\footnotesize{T-NN w/o TL}};
    
    \addplot[color=violet, mark=*, very thick]     
   coordinates {(1,-0.04758)(2,-0.04758)(3,-0.04758)(5,-0.04758)(10,-0.04758)(15,-0.04758)(20,-0.04758)(25,-0.04758)(30,-0.04758)(35,-0.04758)(40,-0.04758)(45,-0.04758)(50,-0.04758)(60,-0.04758)(70,-0.04758)(80,-0.04758)(90,-0.04758)(100,-0.04758)(110,-0.04758)(120,-0.04758)(130,-0.04758)(140,-0.04758)(150,-0.04758)};
    \addlegendentry{\footnotesize{S-NN}};

    \addplot[color=orange, mark=triangle, very thick]     
   coordinates {(1,4.1443)(2,4.1443)(3,4.1443)(5,4.1443)(10,4.1443)(15,4.1443)(20,4.1443)(25,4.1443)(30,4.1443)(35,4.1443)(40,4.1443)(45,4.1443)(50,4.1443)(60,4.1443)(70,4.1443)(80,4.1443)(90,4.1443)(100,4.1443)(110,4.1443)(120,4.1443)(130,4.1443)(140,4.1443)(150,4.1443)};
    \addlegendentry{\footnotesize{w/o NN}};
    
  \end{axis}
    \end{tikzpicture}
  \caption{ }
\end{subfigure}\hfil 
\begin{subfigure}{0.33\textwidth}
  \begin{tikzpicture}[scale=0.65]
    \begin{axis} [
        xlabel={Number of epochs},
        ylabel={Q-Factor [dB]},
        grid=both,          xtick = {1,20,40,60,80,100,120,140},   
    	xmin=1, xmax=150,
        legend style={legend style={ at={(1,0.3)},anchor= east}, legend cell align=left,fill=white, fill opacity=0.6, draw opacity=1,text opacity=1},
    	grid style={dashed}]
        ]
      \addplot[color=red, mark=square, very thick]
    coordinates {(1,5.2232)(2,5.5536)(3,5.676)(5,5.7822)(10,5.8862)(15,5.9101)(20,5.9205)(25,5.9288)(30,5.9386)(35,5.92)(40,5.9262)(45,5.9227)(50,5.9222)(60,5.935)(70,5.9324)(80,5.9417)(90,5.9371)(100,5.9367)(110,5.9249)(120,5.9165)(130,5.9258)(140,5.9366)(150,5.9366)};

    \addlegendentry{\footnotesize{T-NN TL 100\%}};

    \addplot[color=green, mark=diamond, very thick]
  coordinates {(1,3.2364)(2,3.923)(3,4.3282)(5,4.704)(10,5.0267)(15,5.1512)(20,5.2196)(25,5.2697)(30,5.3113)(35,5.338)(40,5.34)(45,5.3579)(50,5.3487)(60,5.3824)(70,5.3936)(80,5.3738)(90,5.4123)(100,5.4096)(110,5.4017)(120,5.4017)(130,5.4017)(140,5.4017)(150,5.4017)};
    \addlegendentry{\footnotesize{T-NN TL 20\%}};
    
    \addplot[color=blue, mark=o, very thick]   
    coordinates {(1,2.242)(2,2.3129)(3,2.3373)(5,2.5716)(10,3.8669)(15,4.2836)(20,4.5908)(25,4.7683)(30,4.9204)(35,5.0201)(40,5.1104)(45,5.2662)(50,5.4095)(60,5.5298)(70,5.5317)(80,5.6624)(90,5.7468)(100,5.7722)(110,5.8441)(120,5.8764)(130,5.9074)(140,5.8252)(150,5.9348)};
    \addlegendentry{\footnotesize{T-NN w/o TL}};
    
    \addplot[color=violet, mark=*, very thick]     
    coordinates {(1,-2.1325)(2,-2.1325)(3,-2.1325)(5,-2.1325)(10,-2.1325)(15,-2.1325)(20,-2.1325)(25,-2.1325)(30,-2.1325)(35,-2.1325)(40,-2.1325)(45,-2.1325)(50,-2.1325)(60,-2.1325)(70,-2.1325)(80,-2.1325)(90,-2.1325)(100,-2.1325)(110,-2.1325)(120,-2.1325)(130,-2.1325)(140,-2.1325)(150,-2.1325)};
    \addlegendentry{\footnotesize{S-NN}};

    \addplot[color=orange, mark=triangle, very thick]  coordinates {(1,2.4823)(2,2.4823)(3,2.4823)(5,2.4823)(10,2.4823)(15,2.4823)(20,2.4823)(25,2.4823)(30,2.4823)(35,2.4823)(40,2.4823)(45,2.4823)(50,2.4823)(60,2.4823)(70,2.4823)(80,2.4823)(90,2.4823)(100,2.4823)(110,2.4823)(120,2.4823)(130,2.4823)(140,2.4823)(150,2.4823)};
    \addlegendentry{\footnotesize{w/o NN}};
  \end{axis}
    \end{tikzpicture}
  \caption{ }
\end{subfigure}

\medskip
\begin{subfigure}{0.33\textwidth}
  \begin{tikzpicture}[scale=0.65]
    \begin{axis} [
        xlabel={Number of epochs},
        ylabel={Q-Factor [dB]},
        grid=both,          xtick = {1,20,40,60,80,100},   
    	xmin=1, xmax=100,
        legend style={legend style={ at={(1,0.3)},anchor= east}, legend cell align=left,fill=white, fill opacity=0.6, draw opacity=1,text opacity=1},
    	grid style={dashed}]
        ]
      \addplot[color=red, mark=square, very thick]
    coordinates {(1,5.6033)(2,6.4258)(3,6.6784)(5,6.8264)(10,6.9328)(15,6.9687)(20,6.9725)(25,6.9729)(30,6.9816)(35,6.975)(41,6.9839)(45,6.9779)(50,6.9758)(55,6.9758)(60,6.9758)(65,6.9758)(70,6.9758)(75,6.9758)(80,6.9758)(85,6.9758)(90,6.9758)(95,6.9758)(100,6.9758)};
    \addlegendentry{\footnotesize{T-NN TL 100\%}};

    \addplot[color=green, mark=diamond, very thick]
  coordinates {(1,2.9627)(2,3.2679)(3,3.7316)(5,4.4198)(10,5.6677)(15,6.2306)(20,6.478)(25,6.6209)(30,6.7103)(35,6.7768)(41,6.8067)(45,6.8234)(50,6.8641)(55,6.8641)(60,6.8641)(65,6.8641)(70,6.8641)(75,6.8641)(80,6.8641)(85,6.8641)(90,6.8641)(95,6.8641)(100,6.8641)};
    \addlegendentry{\footnotesize{T-NN TL 10\%}};
    
    \addplot[color=blue, mark=o, very thick]   
   coordinates {(1,2.774)(2,2.8562)(3,3.3204)(5,5.1136)(10,5.9427)(15,6.3477)(20,6.4597)(25,6.6003)(30,6.6927)(35,6.7282)(41,6.7164)(45,6.7769)(50,6.7532)(55,6.77)(60,6.8208)(65,6.8667)(70,6.8159)(75,6.8537)(80,6.8124)(85,6.8469)(90,6.8327)(95,6.8753)(100,6.8876)};
    \addlegendentry{\footnotesize{T-NN w/o TL}};
    
    \addplot[color=violet, mark=*, very thick]     
    coordinates {(1,1.7493)(2,1.7493)(3,1.7493)(5,1.7493)(10,1.7493)(15,1.7493)(20,1.7493)(25,1.7493)(30,1.7493)(35,1.7493)(41,1.7493)(45,1.7493)(50,1.7493)(55,1.7493)(60,1.7493)(65,1.7493)(70,1.7493)(75,1.7493)(80,1.7493)(85,1.7493)(90,1.7493)(95,1.7493)(100,1.7493)};
    \addlegendentry{\footnotesize{S-NN}};

    \addplot[color=orange, mark=triangle, very thick]  coordinates {(1,3.0156)(2,3.0156)(3,3.0156)(5,3.0156)(10,3.0156)(15,3.0156)(20,3.0156)(25,3.0156)(30,3.0156)(35,3.0156)(41,3.0156)(45,3.0156)(50,3.0156)(55,3.0156)(60,3.0156)(65,3.0156)(70,3.0156)(75,3.0156)(80,3.0156)(85,3.0156)(90,3.0156)(95,3.0156)(100,3.0156)};
    \addlegendentry{\footnotesize{w/o NN}};
  \end{axis}
    \end{tikzpicture}
  \caption{ }
\end{subfigure}\hfil 
\begin{subfigure}{0.33\textwidth}
  \begin{tikzpicture}[scale=0.65]
    \begin{axis} [
        xlabel={Number of epochs},
        ylabel={Q-Factor [dB]},
        grid=both,          xtick = {1,20,40,60,80,100},   
    	xmin=1, xmax=100,
        legend style={legend style={ at={(1,0.3)},anchor= east}, legend cell align=left,fill=white, fill opacity=0.6, draw opacity=1,text opacity=1},
    	grid style={dashed}]
        ]
      \addplot[color=red, mark=square, very thick]
    coordinates {(1,3.6993)(2,4.327)(3,4.4996)(5,4.6082)(10,4.6848)(15,4.6959)(20,4.7158)(25,4.7159)(30,4.7192)(35,4.7179)(41,4.717)(45,4.7173)(50,4.7147)(55,4.7147)(60,4.7147)(65,4.7147)(70,4.7147)(75,4.7147)(80,4.7147)(85,4.7147)(90,4.7147)(95,4.7147)(100,4.7147)};
    \addlegendentry{\footnotesize{T-NN TL 100\%}};

    \addplot[color=green, mark=diamond, very thick]
    coordinates {(1,1.488)(2,1.6733)(3,2.1752)(5,2.7909)(10,3.7781)(15,4.1744)(20,4.3641)(25,4.4599)(30,4.5148)(35,4.5593)(41,4.5699)(45,4.5984)(50,4.6164)(55,4.6164)(60,4.6164)(65,4.6164)(70,4.6164)(75,4.6164)(80,4.6164)(85,4.6164)(90,4.6164)(95,4.6164)(100,4.6164)};
    \addlegendentry{\footnotesize{T-NN TL 10\%}};
    
    \addplot[color=blue, mark=o, very thick]   
    coordinates {(1,1.104)(2,1.269)(3,1.7198)(5,3.1636)(10,3.8704)(15,4.1631)(20,4.3295)(25,4.3108)(30,4.3373)(35,4.4376)(41,4.4888)(45,4.5741)(50,4.5496)(55,4.5515)(60,4.5721)(65,4.6151)(70,4.6004)(75,4.626)(80,4.6122)(85,4.5904)(90,4.6417)(95,4.5912)(100,4.6405)};
    \addlegendentry{\footnotesize{T-NN w/o TL}};
    
    \addplot[color=violet, mark=*, very thick]     
    coordinates {(1,0.57647)(2,0.57647)(3,0.57647)(5,0.57647)(10,0.57647)(15,0.57647)(20,0.57647)(25,0.57647)(30,0.57647)(35,0.57647)(41,0.57647)(45,0.57647)(50,0.57647)(55,0.57647)(60,0.57647)(65,0.57647)(70,0.57647)(75,0.57647)(80,0.57647)(85,0.57647)(90,0.57647)(95,0.57647)(100,0.57647)};
    \addlegendentry{\footnotesize{S-NN}};

    \addplot[color=orange, mark=triangle, very thick]     
    coordinates {(1,1.3282)(2,1.3282)(3,1.3282)(5,1.3282)(10,1.3282)(15,1.3282)(20,1.3282)(25,1.3282)(30,1.3282)(35,1.3282)(41,1.3282)(45,1.3282)(50,1.3282)(55,1.3282)(60,1.3282)(65,1.3282)(70,1.3282)(75,1.3282)(80,1.3282)(85,1.3282)(90,1.3282)(95,1.3282)(100,1.3282)};
    \addlegendentry{\footnotesize{w/o NN}};

  \end{axis}
    \end{tikzpicture}
  \caption{ }
\end{subfigure}\hfil 
\begin{subfigure}{0.33\textwidth}
  \begin{tikzpicture}[scale=0.65]
    \begin{axis} [
        xlabel={Number of epochs},
        ylabel={Q-Factor [dB]},
        grid=both,          xtick = {1,20,40,60,80,100},   
    	xmin=1, xmax=100,
        legend style={legend style={ at={(1,0.3)},anchor= east}, legend cell align=left,fill=white, fill opacity=0.6, draw opacity=1,text opacity=1},
    	grid style={dashed}]
        ]
      \addplot[color=red, mark=square, very thick]
     coordinates {(1,1.8467)(2,2.4609)(3,2.6252)(5,2.7388)(10,2.8006)(15,2.8192)(20,2.8329)(25,2.8317)(30,2.8279)(35,2.842)(41,2.8346)(45,2.8392)(50,2.8371)(55,2.8371)(60,2.8371)(65,2.8371)(70,2.8371)(75,2.8371)(80,2.8371)(85,2.8371)(90,2.8371)(95,2.8371)(100,2.8371)};
    \addlegendentry{\footnotesize{T-NN TL 100\%}};

    \addplot[color=green, mark=diamond, very thick]
  coordinates {(1,-0.23032)(2,-0.069854)(3,0.43405)(5,0.99085)(10,1.9176)(15,2.3071)(20,2.514)(25,2.5944)(30,2.6514)(35,2.6927)(41,2.7263)(45,2.7529)(50,2.7553)(55,2.7553)(60,2.7553)(65,2.7553)(70,2.7553)(75,2.7553)(80,2.7553)(85,2.7553)(90,2.7553)(95,2.7553)(100,2.7553)};
    \addlegendentry{\footnotesize{T-NN TL 10\%}};
    
    \addplot[color=blue, mark=o, very thick]   
    coordinates {(1,-0.46653)(2,-0.35225)(3,0.032936)(5,1.4213)(10,2.1435)(15,2.3663)(20,2.5351)(25,2.5695)(30,2.6371)(35,2.6855)(41,2.6829)(45,2.6947)(50,2.6812)(55,2.6954)(60,2.7369)(65,2.718)(70,2.6822)(75,2.7122)(80,2.7607)(85,2.7617)(90,2.7853)(95,2.7503)(100,2.7604)};
    \addlegendentry{\footnotesize{T-NN w/o TL}};
    
    \addplot[color=violet, mark=*, very thick]     
    coordinates {(1,-1.287)(2,-1.287)(3,-1.287)(5,-1.287)(10,-1.287)(15,-1.287)(20,-1.287)(25,-1.287)(30,-1.287)(35,-1.287)(41,-1.287)(45,-1.287)(50,-1.287)(55,-1.287)(60,-1.287)(65,-1.287)(70,-1.287)(75,-1.287)(80,-1.287)(85,-1.287)(90,-1.287)(95,-1.287)(100,-1.287)};
    \addlegendentry{\footnotesize{S-NN}};

    \addplot[color=orange, mark=triangle, very thick]  
    coordinates {(1,-0.24089)(2,-0.24089)(3,-0.24089)(5,-0.24089)(10,-0.24089)(15,-0.24089)(20,-0.24089)(25,-0.24089)(30,-0.24089)(35,-0.24089)(41,-0.24089)(45,-0.24089)(50,-0.24089)(55,-0.24089)(60,-0.24089)(65,-0.24089)(70,-0.24089)(75,-0.24089)(80,-0.24089)(85,-0.24089)(90,-0.24089)(95,-0.24089)(100,-0.24089)};
    \addlegendentry{\footnotesize{w/o NN}};
  \end{axis}
    \end{tikzpicture}
  \caption{ }
\end{subfigure}
\caption{Transferring the learning between modulation formats. Case I: from 16-QAM to (a) 32-QAM, (b) 64-QAM, (c) 128-QAM, using 18$\times$50~km SSMF fiber link and 4~dBm 34.4~GBd signals. Case II: from 16-QAM to (d) 32-QAM, (e) 64-QAM, (f) 128-QAM, using 9$\times$50~km TWC fiber link and 2~dBm at 34.4~GBd signals. Case III: from 8~dBm / 16-QAM to (g) 4~dBm / 32-QAM, (h) 4~dBm / 64-QAM, (i) 4~dBm / 128-QAM, using 18$\times$50~km SSMF fiber link and 34.4~GBd. Case IV: 5~dBm / 16-QAM to (j) 2~dBm / 32-QAM, (k) 2~dBm / 64-QAM, (l) 2~dBm / 128-QAM, using 9$\times$50~km TWC fiber link and 34.4~GBd.}
\label{fig:transfer_MF}
\end{figure*}

\subsection{Transfer learning for different modulation formats}\label{subSec:MF}

In this section, we analyze the impact of changing the modulation format on the NN equalizer's performance. The source dataset modulation format is 16-QAM, whereas the target modulation formats are 32-QAM, 64-QAM, and 128-QAM. First, the same launch power is kept independently of the selected modulation format. Fig.~\ref{fig:transfer_MF} shows the results for the two cases studied where only the modulation format changes: Case I (Fig.~\ref{fig:transfer_MF}-a, b, c) with the SSMF setup at a launch power of 4~dBm, and Case II (Fig.~\ref{fig:transfer_MF}-d, e, f) with the TWC setup at a launch power of 2~dBm. 


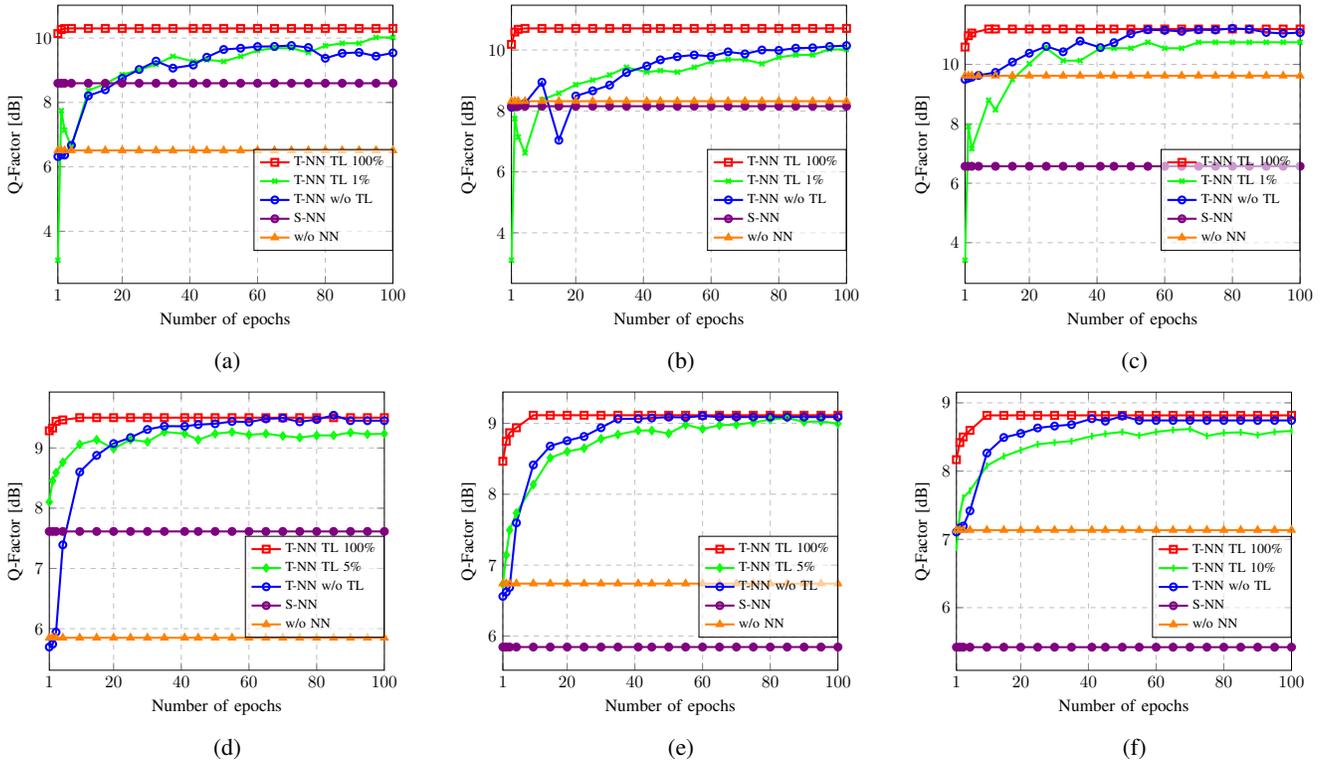
\begin{figure*}[ht!]
    \centering 
\begin{subfigure}{0.33\textwidth}
  \begin{tikzpicture}[scale=0.65]
    \begin{axis} [
        xlabel={Number of epochs},
        ylabel={Q-Factor [dB]},
        grid=both,          xtick = {1,20,40,60,80,100},   
    	xmin=1, xmax=100,
        legend style={legend style={ at={(1,0.3)},anchor= east}, legend cell align=left,fill=white, fill opacity=0.6, draw opacity=1,text opacity=1},
    	grid style={dashed}]
        ]
      \addplot[color=red, mark=square, very thick]
    coordinates {(1,10.139)(2,10.2658)(3,10.2961)(5,10.3001)(10,10.3001)(15,10.3001)(20,10.3001)(25,10.3001)(30,10.3001)(35,10.3001)(41,10.3001)(45,10.3001)(50,10.3001)(55,10.3001)(60,10.3001)(65,10.3001)(70,10.3001)(75,10.3001)(80,10.3001)(85,10.3001)(90,10.3001)(95,10.3001)(100,10.3001)};
    \addlegendentry{\footnotesize{T-NN TL 100\%}};
    
    \addplot[color=green, mark=x, very thick]
  coordinates {(1,3.1008)(2,7.7476)(3,7.1405)(5,6.6237)(10,8.3683)(15,8.5795)(20,8.863)(25,9.0118)(30,9.1809)(35,9.4329)(41,9.2752)(45,9.3254)(50,9.2752)(55,9.4329)(60,9.6167)(65,9.6857)(70,9.6857)(75,9.552)(80,9.7595)(85,9.8391)(90,9.8391)(95,10.0203)(100,10.0203)};
    \addlegendentry{\footnotesize{T-NN TL 1\%}};
    
    \addplot[color=blue, mark=o, very thick]   
    coordinates {(1,6.3195)(2,6.3662)(3,6.3672)(5,6.6696)(10,8.207)(15,8.3943)(20,8.7493)(25,9.0239)(30,9.2842)(35,9.0669)(41,9.1626)(45,9.4047)(50,9.6453)(55,9.6837)(60,9.7362)(65,9.7415)(70,9.7689)(75,9.7067)(80,9.3732)(85,9.5291)(90,9.5546)(95,9.4382)(100,9.5409)};
    \addlegendentry{\footnotesize{T-NN w/o TL}};
    
    \addplot[color=violet, mark=*, very thick]     
   coordinates {(1,8.5978)(2,8.5978)(3,8.5978)(5,8.5978)(10,8.5978)(15,8.5978)(20,8.5978)(25,8.5978)(30,8.5978)(35,8.5978)(41,8.5978)(45,8.5978)(50,8.5978)(55,8.5978)(60,8.5978)(65,8.5978)(70,8.5978)(75,8.5978)(80,8.5978)(85,8.5978)(90,8.5978)(95,8.5978)(100,8.5978)};
    \addlegendentry{\footnotesize{S-NN}};

    \addplot[color=orange, mark=triangle, very thick]     
    coordinates {(1,6.5102)(2,6.5102)(3,6.5102)(5,6.5102)(10,6.5102)(15,6.5102)(20,6.5102)(25,6.5102)(30,6.5102)(35,6.5102)(41,6.5102)(45,6.5102)(50,6.5102)(55,6.5102)(60,6.5102)(65,6.5102)(70,6.5102)(75,6.5102)(80,6.5102)(85,6.5102)(90,6.5102)(95,6.5102)(100,6.5102)};
    \addlegendentry{\footnotesize{w/o NN}};
  \end{axis}
    \end{tikzpicture}
  \caption{ }
\end{subfigure}\hfil 
\begin{subfigure}{0.33\textwidth}
  \begin{tikzpicture}[scale=0.65]
    \begin{axis} [
        xlabel={Number of epochs},
        ylabel={Q-Factor [dB]},
        grid=both,          xtick = {1,20,40,60,80,100},   
    	xmin=1, xmax=100,
        legend style={legend style={ at={(1,0.3)},anchor= east}, legend cell align=left,fill=white, fill opacity=0.6, draw opacity=1,text opacity=1},
    	grid style={dashed}]
        ]
      \addplot[color=red, mark=square, very thick]
    coordinates {(1,10.1864)(2,10.5896)(3,10.6791)(5,10.709)(10,10.709)(15,10.709)(20,10.709)(25,10.709)(30,10.709)(35,10.709)(41,10.709)(45,10.709)(50,10.709)(55,10.709)(60,10.709)(65,10.709)(70,10.709)(75,10.709)(80,10.709)(85,10.709)(90,10.709)(95,10.709)(100,10.709)};
    \addlegendentry{\footnotesize{T-NN TL 100\%}};
    
    \addplot[color=green, mark=x, very thick]
  coordinates {(1,3.1008)(2,7.7476)(3,7.1405)(5,6.6237)(10,8.3683)(15,8.5795)(20,8.863)(25,9.0118)(30,9.1809)(35,9.4329)(41,9.2752)(45,9.3254)(50,9.2752)(55,9.4329)(60,9.6167)(65,9.6857)(70,9.6857)(75,9.552)(80,9.7595)(85,9.8391)(90,9.8391)(95,10.0203)(100,10.0203)};
    \addlegendentry{\footnotesize{T-NN TL 1\%}};
    
    \addplot[color=blue, mark=o, very thick]   
    coordinates {(1,8.123)(2,8.1579)(3,8.1706)(5,8.2534)(10,8.9429)(15,7.0388)(20,8.492)(25,8.6611)(30,8.8492)(35,9.2598)(41,9.4774)(45,9.6773)(50,9.7813)(55,9.8376)(60,9.7931)(65,9.9384)(70,9.8721)(75,10.0007)(80,9.9881)(85,10.0611)(90,10.0726)(95,10.1178)(100,10.1446)};
    \addlegendentry{\footnotesize{T-NN w/o TL}};
    
    \addplot[color=violet, mark=*, very thick]     
    coordinates {(1,8.155)(2,8.155)(3,8.155)(5,8.155)(10,8.155)(15,8.155)(20,8.155)(25,8.155)(30,8.155)(35,8.155)(41,8.155)(45,8.155)(50,8.155)(55,8.155)(60,8.155)(65,8.155)(70,8.155)(75,8.155)(80,8.155)(85,8.155)(90,8.155)(95,8.155)(100,8.155)};
    \addlegendentry{\footnotesize{S-NN}};

    \addplot[color=orange, mark=triangle, very thick]     
    coordinates {(1,8.3204)(2,8.3204)(3,8.3204)(5,8.3204)(10,8.3204)(15,8.3204)(20,8.3204)(25,8.3204)(30,8.3204)(35,8.3204)(41,8.3204)(45,8.3204)(50,8.3204)(55,8.3204)(60,8.3204)(65,8.3204)(70,8.3204)(75,8.3204)(80,8.3204)(85,8.3204)(90,8.3204)(95,8.3204)(100,8.3204)};
    \addlegendentry{\footnotesize{w/o NN}};
  \end{axis}
    \end{tikzpicture}
  \caption{ }
\end{subfigure}\hfil 
\begin{subfigure}{0.33\textwidth}
  \begin{tikzpicture}[scale=0.65]
    \begin{axis} [
        xlabel={Number of epochs},
        ylabel={Q-Factor [dB]},
        grid=both,          xtick = {1,20,40,60,80,100},   
    	xmin=1, xmax=100,
        legend style={legend style={ at={(1,0.3)},anchor= east}, legend cell align=left,fill=white, fill opacity=0.6, draw opacity=1,text opacity=1},
    	grid style={dashed}]
        ]
      \addplot[color=red, mark=square, very thick]
    coordinates {(1,10.5821)(2,10.9641)(3,11.0539)(8,11.1882)(10,11.1882)(15,11.1882)(20,11.1882)(25,11.1882)(30,11.1882)(35,11.1882)(41,11.1882)(45,11.1882)(50,11.1882)(55,11.1882)(60,11.1882)(65,11.1882)(70,11.1882)(75,11.1882)(80,11.1882)(85,11.1882)(90,11.1882)(95,11.1882)(100,11.1882)};
    \addlegendentry{\footnotesize{T-NN TL 100\%}};
    
    \addplot[color=green, mark=x, very thick]
  coordinates {(1,3.4093)(2,7.9049)(3,7.1649)(8,8.7946)(10,8.47)(15,9.4908)(20,10.0203)(25,10.5419)(30,10.1254)(35,10.1254)(41,10.5419)(45,10.5419)(50,10.5419)(55,10.7423)(60,10.5419)(65,10.5419)(70,10.7423)(75,10.7423)(80,10.7423)(85,10.7423)(90,10.7423)(95,10.7423)(100,10.7423)};
    \addlegendentry{\footnotesize{T-NN TL 1\%}};
    \addplot[color=blue, mark=o, very thick]   
  coordinates {(1,9.4946)(2,9.5334)(3,9.5673)(5,9.6296)(10,9.728)(15,10.0778)(20,10.3745)(25,10.6068)(30,10.4227)(35,10.782)(41,10.5637)(45,10.7289)(50,11.0271)(55,11.164)(60,11.1445)(65,11.1106)(70,11.164)(75,11.1561)(80,11.209)(85,11.1801)(90,11.0746)(95,11.0404)(100,11.0677)};
    \addlegendentry{\footnotesize{T-NN w/o TL}};
    
    \addplot[color=violet, mark=*, very thick]     
  coordinates {(1,6.57)(2,6.57)(3,6.57)(5,6.57)(10,6.57)(15,6.57)(20,6.57)(25,6.57)(30,6.57)(35,6.57)(41,6.57)(45,6.57)(50,6.57)(55,6.57)(60,6.57)(65,6.57)(70,6.57)(75,6.57)(80,6.57)(85,6.57)(90,6.57)(95,6.57)(100,6.57)};
    \addlegendentry{\footnotesize{S-NN}};

    \addplot[color=orange, mark=triangle, very thick]     
    coordinates {(1,9.6167)(2,9.6167)(3,9.6167)(5,9.6167)(10,9.6167)(15,9.6167)(20,9.6167)(25,9.6167)(30,9.6167)(35,9.6167)(41,9.6167)(45,9.6167)(50,9.6167)(55,9.6167)(60,9.6167)(65,9.6167)(70,9.6167)(75,9.6167)(80,9.6167)(85,9.6167)(90,9.6167)(95,9.6167)(100,9.6167)};
    \addlegendentry{\footnotesize{w/o NN}};
  \end{axis}
    \end{tikzpicture}
  \caption{ }
\end{subfigure}

\medskip
\begin{subfigure}{0.33\textwidth}
  \begin{tikzpicture}[scale=0.65]
    \begin{axis} [
        xlabel={Number of epochs},
        ylabel={Q-Factor [dB]},
        grid=both,          xtick = {1,20,40,60,80,100},   
    	xmin=1, xmax=100,
        legend style={legend style={ at={(1,0.3)},anchor= east}, legend cell align=left,fill=white, fill opacity=0.6, draw opacity=1,text opacity=1},
    	grid style={dashed}]
        ]
      \addplot[color=red, mark=square, very thick]
    coordinates {(1,9.2847)(2,9.3302)(3,9.4416)(5,9.4657)(10,9.5029)(15,9.5029)(20,9.5029)(25,9.5029)(30,9.5029)(35,9.5029)(41,9.5029)(45,9.5029)(50,9.5029)(55,9.5029)(60,9.5029)(65,9.5029)(70,9.5029)(75,9.5029)(80,9.5029)(85,9.5029)(90,9.5029)(95,9.5029)(100,9.5029)};
    \addlegendentry{\footnotesize{T-NN TL 100\%}};

    \addplot[color=green, mark=diamond, very thick]
  coordinates {(1,8.1032)(2,8.4545)(3,8.591)(5,8.7619)(10,9.0603)(15,9.1365)(20,8.9885)(25,9.1365)(30,9.102)(35,9.2656)(41,9.2367)(45,9.1365)(50,9.2367)(55,9.2656)(60,9.2178)(65,9.2367)(70,9.1993)(75,9.172)(80,9.2085)(85,9.2085)(90,9.2559)(95,9.2272)(100,9.2367)};
    \addlegendentry{\footnotesize{T-NN TL 5\%}};
    
    \addplot[color=blue, mark=o, very thick]   
    coordinates {(1,5.6951)(2,5.7431)(3,5.9429)(5,7.3915)(10,8.6014)(15,8.8766)(20,9.0731)(25,9.1724)(30,9.3067)(35,9.3621)(41,9.3599)(45,9.3878)(50,9.4014)(55,9.4393)(60,9.4292)(65,9.4833)(70,9.4946)(75,9.4337)(80,9.4721)(85,9.5433)(90,9.453)(95,9.453)(100,9.453)};
    \addlegendentry{\footnotesize{T-NN w/o TL}};
    
    \addplot[color=violet, mark=*, very thick]     
    coordinates {(1,7.614)(2,7.614)(3,7.614)(5,7.614)(10,7.614)(15,7.614)(20,7.614)(25,7.614)(30,7.614)(35,7.614)(41,7.614)(45,7.614)(50,7.614)(55,7.614)(60,7.614)(65,7.614)(70,7.614)(75,7.614)(80,7.614)(85,7.614)(90,7.614)(95,7.614)(100,7.614)};
    \addlegendentry{\footnotesize{S-NN}};

    \addplot[color=orange, mark=triangle, very thick] 
    coordinates {(1,5.8488)(2,5.8488)(3,5.8488)(5,5.8488)(10,5.8488)(15,5.8488)(20,5.8488)(25,5.8488)(30,5.8488)(35,5.8488)(41,5.8488)(45,5.8488)(50,5.8488)(55,5.8488)(60,5.8488)(65,5.8488)(70,5.8488)(75,5.8488)(80,5.8488)(85,5.8488)(90,5.8488)(95,5.8488)(100,5.8488)};
    \addlegendentry{\footnotesize{w/o NN}};
  \end{axis}
    \end{tikzpicture}
  \caption{ }
\end{subfigure}\hfil 
\begin{subfigure}{0.33\textwidth}
  \begin{tikzpicture}[scale=0.65]
    \begin{axis} [
        xlabel={Number of epochs},
        ylabel={Q-Factor [dB]},
        grid=both,          xtick = {1,20,40,60,80,100},   
    	xmin=1, xmax=100,
        legend style={legend style={ at={(1,0.3)},anchor= east}, legend cell align=left,fill=white, fill opacity=0.6, draw opacity=1,text opacity=1},
    	grid style={dashed}]
        ]
      \addplot[color=red, mark=square, very thick]
    coordinates {(1,8.4652)(2,8.7464)(3,8.8678)(5,8.9373)(10,9.1142)(15,9.1142)(20,9.1142)(25,9.1142)(30,9.1142)(35,9.1142)(41,9.1142)(45,9.1142)(50,9.1142)(55,9.1142)(60,9.1142)(65,9.1142)(70,9.1142)(75,9.1142)(80,9.1142)(85,9.1142)(90,9.1142)(95,9.1142)(100,9.1142)};
    \addlegendentry{\footnotesize{T-NN TL 100\%}};

    \addplot[color=green, mark=diamond, very thick]
    coordinates {(1,6.7304)(2,7.1407)(3,7.4929)(5,7.7345)(10,8.1366)(15,8.513)(20,8.6026)(25,8.6496)(30,8.7815)(35,8.8423)(41,8.8988)(45,8.8988)(50,8.8562)(55,8.9808)(60,8.9206)(65,8.9731)(70,8.9808)(75,9.012)(80,9.0521)(85,9.0768)(90,9.0199)(95,9.0279)(100,8.9963)};
    \addlegendentry{\footnotesize{T-NN TL 5\%}};
    \addplot[color=blue, mark=o, very thick]   
    coordinates {(1,6.5597)(2,6.619)(3,6.6862)(5,7.5959)(10,8.4129)(15,8.6787)(20,8.7547)(25,8.8149)(30,8.938)(35,9.0628)(41,9.0644)(45,9.0756)(50,9.0877)(55,9.0827)(60,9.1042)(65,9.089)(70,9.089)(75,9.089)(80,9.089)(85,9.089)(90,9.089)(95,9.089)(100,9.089)};
    \addlegendentry{\footnotesize{T-NN w/o TL}};
    
    \addplot[color=violet, mark=*, very thick]     
    coordinates {(1,5.8446)(2,5.8446)(3,5.8446)(5,5.8446)(10,5.8446)(15,5.8446)(20,5.8446)(25,5.8446)(30,5.8446)(35,5.8446)(41,5.8446)(45,5.8446)(50,5.8446)(55,5.8446)(60,5.8446)(65,5.8446)(70,5.8446)(75,5.8446)(80,5.8446)(85,5.8446)(90,5.8446)(95,5.8446)(100,5.8446)};
    \addlegendentry{\footnotesize{S-NN}};

    \addplot[color=orange, mark=triangle, very thick]     
    coordinates {(1,6.7391)(2,6.7391)(3,6.7391)(5,6.7391)(10,6.7391)(15,6.7391)(20,6.7391)(25,6.7391)(30,6.7391)(35,6.7391)(41,6.7391)(45,6.7391)(50,6.7391)(55,6.7391)(60,6.7391)(65,6.7391)(70,6.7391)(75,6.7391)(80,6.7391)(85,6.7391)(90,6.7391)(95,6.7391)(100,6.7391)};
    \addlegendentry{\footnotesize{w/o NN}};
  \end{axis}
    \end{tikzpicture}
  \caption{ }
\end{subfigure}\hfil 
\begin{subfigure}{0.33\textwidth}
  \begin{tikzpicture}[scale=0.65]
    \begin{axis} [
        xlabel={Number of epochs},
        ylabel={Q-Factor [dB]},
        grid=both,          xtick = {1,20,40,60,80,100},   
    	xmin=1, xmax=100,
        legend style={legend style={ at={(1,0.3)},anchor= east}, legend cell align=left,fill=white, fill opacity=0.6, draw opacity=1,text opacity=1},
    	grid style={dashed}]
        ]
      \addplot[color=red, mark=square, very thick]
    coordinates {(1,8.1678)(2,8.4174)(3,8.5005)(5,8.5968)(10,8.8165)(15,8.8165)(20,8.8165)(25,8.8165)(30,8.8165)(35,8.8165)(41,8.8165)(45,8.8165)(50,8.8165)(55,8.8165)(60,8.8165)(65,8.8165)(70,8.8165)(75,8.8165)(80,8.8165)(85,8.8165)(90,8.8165)(95,8.8165)(100,8.8165)};
    \addlegendentry{\footnotesize{T-NN TL 100\%}};
    
    \addplot[color=green, mark=|, very thick]
  coordinates {(1,6.8793)(2,7.3679)(3,7.6097)(5,7.7131)(10,8.0786)(15,8.2167)(20,8.3083)(25,8.3932)(30,8.4184)(35,8.4389)(41,8.513)(45,8.5459)(50,8.5739)(55,8.5211)(60,8.5739)(65,8.6026)(70,8.6171)(75,8.5157)(80,8.5598)(85,8.5683)(90,8.5293)(95,8.5739)(100,8.5882)};
    \addlegendentry{\footnotesize{T-NN TL 10\%}};
    
    \addplot[color=blue, mark=o, very thick]   
    coordinates {(1,7.1087)(2,7.1689)(3,7.1948)(5,7.4184)(10,8.2639)(15,8.4922)(20,8.5515)(25,8.6321)(30,8.6602)(35,8.6811)(41,8.7674)(45,8.732)(50,8.8065)(55,8.7425)(60,8.7425)(65,8.7425)(70,8.7425)(75,8.7425)(80,8.7425)(85,8.7425)(90,8.7425)(95,8.7425)(100,8.7425)};
    \addlegendentry{\footnotesize{T-NN w/o TL}};
    
    \addplot[color=violet, mark=*, very thick]     
    coordinates {(1,5.4213)(2,5.4213)(3,5.4213)(5,5.4213)(10,5.4213)(15,5.4213)(20,5.4213)(25,5.4213)(30,5.4213)(35,5.4213)(41,5.4213)(45,5.4213)(50,5.4213)(55,5.4213)(60,5.4213)(65,5.4213)(70,5.4213)(75,5.4213)(80,5.4213)(85,5.4213)(90,5.4213)(95,5.4213)(100,5.4213)};

    \addlegendentry{\footnotesize{S-NN}};

    \addplot[color=orange, mark=triangle, very thick]     
    coordinates {(1,7.1359)(2,7.1359)(3,7.1359)(5,7.1359)(10,7.1359)(15,7.1359)(20,7.1359)(25,7.1359)(30,7.1359)(35,7.1359)(41,7.1359)(45,7.1359)(50,7.1359)(55,7.1359)(60,7.1359)(65,7.1359)(70,7.1359)(75,7.1359)(80,7.1359)(85,7.1359)(90,7.1359)(95,7.1359)(100,7.1359)};
    \addlegendentry{\footnotesize{w/o NN}};
  \end{axis}
    \end{tikzpicture}
  \caption{ }
\end{subfigure}
\caption{Transfer learning between symbol rates. Case I: from 34.4~GBd to (a) 45~GBd, (b) 65~GBd, (c) 85~GBd, in a 18$\times$50~km SSMF link using 16-QAM and 6~dBm of launch power. Case II: from 34.4~GBd to (d) 45~GBd, (e) 65~GBd, (f) 85~GBd, in a 9$\times$50~km TWC fiber link using 16-QAM and 2~dBm of launch power.}
\label{fig:transfer_SR}
\end{figure*}

In this scenario, only the convolutional layers were retrained in the case of T-NN with TL. From the results obtained, we can infer that both the T-NN with TL and S-NN can be successfully used with different modulation formats, as can be readily seen from Fig.~\ref{fig:transfer_MF}. This means that we do not have to retrain the model if we change only the constellation size of the modulation format. These results also demonstrate that the nonlinear propagation channel law is almost unaffected by changing the modulation format (from the 16-QAM to a higher-order one) when the power level remains the same. Thus, the NN equalizer, which is reverting the channel nonlinear effects, continues to function well for other modulation formats. This is in stark contrast to the case of classification equalizers (classifiers)~\cite{deligiannidis2020compensation, schaedlerrecurrent, liu2021bi}, because the latter incorporates the decision boundaries in the NN structure itself. For the classifiers, the S-NN will not work with the new target task since its output stage does not capture the different symbol alphabet. For the NN structure used in this work (the CNN+biLSTM with regression), the channel reversion capability of the equalizer is independent of the modulation format. In contrast, for the classification task, the number of neurons in the last layer is defined by the number of constellation points. So, for the classification model, the output layer with an updated dimensionality should be retrained to correctly identify the new constellation points. We note that the small performance deviation of the T-NN with and without TL shown in Fig.~\ref{fig:transfer_MF}, is a consequence of the particular weight initialization. Finally, we note that the TL direction in the case of the modulation format modification is irrelevant to the TL performance inasmuch as the regression-based NN functionality is not affected by the format changes.

The performance of TL when we simultaneously change the modulation format and launch power is also evaluated in Fig.~\ref{fig:transfer_MF}. The source dataset in Case III was the transmission of 16-QAM signals with a launch power of 8~dBm in the SSMF link, and we transfer the learned parameters to the target having 4~dBm launch power and (g) 32-QAM, (h) 64-QAM, (i) 128-QAM modulation formats. The source dataset of Case IV was the transmission of 16-QAM signals with a launch power of 5~dBm in the TWC fiber link, and we transfer the learned parameters to the target with 2~dBm launch power and (j) 32-QAM, (k) 64-QAM, (l) 128-QAM modulation formats. From the analysis of Cases III and IV in Fig.~\ref{fig:transfer_MF}, we can notice a reduction of up to 95\% in epochs and 90\% in the training dataset for the SSMF case, and the decrease of up to 85\% in epochs and 90\% in the training dataset for the TWC case. As expected, these figures are close to the ones obtained when evaluating the TL between different launch powers, Subsec.~\ref{subSec:power}.

\subsection{Transfer learning for different symbol rates}\label{subSec:symbolrate}
In this section, we evaluate the functionality of TL when only the symbol rate is changed. By changing the symbol rate and keeping the remaining system parameters constant, we effectively change how the neighboring symbols interact with each other. In other words, this change impacts the channel memory. As the channel memory is primarily handled by the biLSTM part of our CNN+biLSTM equalizer, in this section we use the TL defined in Fig.~\ref{fig:TRANSFERLEARNING}, middle panel, inset (b). We retrain the biLSTM and output weights but keep the convolutional weights frozen. Note that when we change the symbol rate, the effective nonlinearity level is also affected. Nonetheless, we observed that we do not need to retrain the CNN layer because the features that were extracted by it (embedded in the values of its kernels), do represent the case of higher nonlinearity, and those can be fine-tuned to the lower nonlinearity by just adjusting the LSTM layer input weights. On the other hand, the weights between the LSTM cells, which cater for the memory effects among other possible representations, had to be retrained because the memory effect changes when the symbol rate does. Our source dataset is the 34.4~GBd signal, and the target symbol rates are 45~GBd, 65~GBd, and 85~GBd. We consider the SSMF and TWC link cases with 16-QAM modulation format and 6~dBm and 2~dBm launch powers, respectively. Fig.~\ref{fig:transfer_SR} depicts the results for the SSMF and TWC fiber links, referred to as Case I and Case II, respectively. The analysis of this figure shows that, by significantly changing the symbol rate with respect to the source, the S-NN performance can degrade below the reference system (w/o NN).

In Case I, when moving to (a) 45~GBd, (b) 65~GBd, and (c) 85~GBd, the number of necessary epochs decreased by 99\%, 95\%, and 81\%, respectively, for the SSMF link case. Furthermore, the re-training process needs much less data, with a reduction of up to 99\% of the required training data for the three symbol rates considered. The number of required epochs for the TWC fiber link (Case II), decreased by 92\%, 73\%, and 75\%, respectively, when switching to (d) 45~GBd, (e) 65~GBd, and (f) 85~GBd. We can also see that the retraining phase requires fewer data: 95\%, 95\%, and 90\%, respectively.
This demonstrates the potential of TL when adjusting the NN to different symbol rates, which is a very important feature when considering the current commercial transponders which can operate in a very wide range of symbol rates.  We point out here that in some cases when retraining with the least possible training data percentage (green curve), a negative transfer (slight performance degradation) occurs at early epochs. This is a situation whereby the randomly selected portion of the training data has a distribution that deviates from the distribution of test data. But we can see that this negative transfer is resolved after just a few epochs (typically less than 10).
This effect is attributed to the difficulty in training recurrent layers \cite{pascanu2013difficulty}. For example, when we reduce the training dataset to 1\% of its original size,  this corresponds to updating our weights approximately 99\% less time per epoch compared to the training with the full dataset. Having such a small amount of updates per epoch for the RNN can lead to instability in the training. But, again, this effect can be sorted out by using several epochs.

\begin{table*}[!hb]
\renewcommand\thetable{III}
\centering
\caption{Summary of the TL effectiveness for the different scenario changes addressed. Rows Test 1 to 6 depict the main results from subsections \ref{subSec:power} (the launch power change), \ref{subSec:MF} (the modulation format change), and \ref{subSec:symbolrate} (the symbol rate change). Rows Test 7 to 12 highlight the results of subsection \ref{subSec:difffiber}, where multiple changes in the transmission configuration, including the change of the fiber type, are analyzed. For computing the epochs saving ratio with the TL, we used 100\% of the training dataset. The red color highlights the particular changes in each test case.}
\begin{tabular}{|c|c|c|c|c|c|c|c|}
\hline
\multirow{2}{*}{\textbf{Test}} & \multicolumn{4}{c|}{\textbf{Scenarios of  TL}}                                              & \multicolumn{3}{c|}{\textbf{Evaluation}}                                      \\ \cline{2-8} 
                               & Fiber                   & Power [dBm]             & Symbol rate [GBd]         & Mod. Format [QAM]       & Max Q-factor[dB]    & Epochs Saved w TL   & Dataset Saved w TL  \\ \hline \hline
1                              & TWC $\rightarrow$ TWC  & {\color[HTML]{FE0000} $5 \rightarrow 3 $}  & $34.4 \rightarrow 34.4 $ & $16 \rightarrow 16 $ & $ 12.66$ & $ 90 \%$ &  $ 94 \%$                   \\ \hline

2                             & SSMF $\rightarrow$ SSMF  & {\color[HTML]{FE0000} $8 \rightarrow 6 $} & $34.4 \rightarrow 34.4 $ & $16 \rightarrow  16 $ & $ 10.16$ & $ 94 \%$ &  $ 94 \%$                   \\ \hline

3                              & TWC $\rightarrow$ TWC  & {\color[HTML]{FE0000}$5 \rightarrow 2 $} & $34.4 \rightarrow 34.4 $ & {\color[HTML]{FE0000}$16 \rightarrow  64 $} & $ 4.64$ & $ 84 \%$ &  $ 90 \%$                   \\ \hline

4                             & SSMF $\rightarrow$ SSMF  & {\color[HTML]{FE0000}$8 \rightarrow 4 $} & $34.4 \rightarrow 34.4 $ & {\color[HTML]{FE0000}$16 \rightarrow  64 $} & $ 8.14$ & $ 94 \%$ &  $ 80 \%$                    \\ \hline

5                             & TWC $\rightarrow$ TWC  & $2 \rightarrow 2 $ & {\color[HTML]{FE0000}$34.4 \rightarrow 45 $} & $16 \rightarrow  16 $ & $ 9.54$ & $ 90 \%$ &  $ 94 \%$                    \\ \hline

6                             & SSMF $\rightarrow$ SSMF  & $6 \rightarrow 6 $ & {\color[HTML]{FE0000}$34.4 \rightarrow 45 $} & $16 \rightarrow  16 $ & $ 9.76$ & $ 98 \%$ &  $ 98 \%$                     \\ \hline \hline

7                              & TWC $\rightarrow$ TWC & {\color[HTML]{FE0000}$5 \rightarrow 3 $} & {\color[HTML]{FE0000}$34.4 \rightarrow 65 $} & {\color[HTML]{FE0000}$16 \rightarrow  64 $} & $ 8.66$ & $ 30\%$ &  $ 50 \%$ \\ \hline

8                              & TWC $\rightarrow$ TWC & {\color[HTML]{FE0000}$8 \rightarrow 4 $} & {\color[HTML]{FE0000}$65 \rightarrow 34.4 $} & $16 \rightarrow  16 $ & $ 10.08$ &  $78\%$ &  $ 90\%$ \\ \hline

9                              & SSMF $\rightarrow$ SSMF & {\color[HTML]{FE0000}$8 \rightarrow 6 $} & {\color[HTML]{FE0000}$34.4 \rightarrow 65 $} & {\color[HTML]{FE0000}$16 \rightarrow  64 $} & $ 5.75$ &  $ 55\%$ &  $ 50 \%$ \\ \hline

10                              & SSMF $\rightarrow$ SSMF & {\color[HTML]{FE0000}$4 \rightarrow 8 $} & {\color[HTML]{FE0000}$34.4 \rightarrow 65 $} & $16 \rightarrow  16 $ & $ 7.24$ &  $ 94\%$ &   $ 99 \%$ \\ \hline

11                              & {\color[HTML]{FE0000}TWC $\rightarrow$ SSMF}  & $5 \rightarrow 5 $ & $34.4 \rightarrow 34.4 $ & $16 \rightarrow  16 $ & $ 11.49$ & $ 10\%$ & $ 50 \%$ \\ \hline

12                              & {\color[HTML]{FE0000}SSMF $\rightarrow$ TWC}  & $5 \rightarrow 5 $ & $34.4 \rightarrow 34.4 $ & $16 \rightarrow  16 $ & $ 10.73$ & $ 0\%$ &  $ 50 \%$ \\ \hline
\end{tabular}
\label{Main_table}
\end{table*}


\begin{table}[htbp]
\renewcommand\thetable{II}
  \centering
  \caption{Dependence of the TL performance on the transfer direction, for the case where we change only the symbol rates from the source to target datasets.}
\begin{tabular}{|c|c|c|c|}
\hline
Fiber                 & Scenario         & Max Q-factor & Epochs required \\ \hline\hline
\multirow{3}{*}{SSMF} & TL 34.4GBd $\rightarrow$ 45GBd  &        10.30      &    \textless 5    \\ \cline{2-4} 
                      & TL 64GBd $\rightarrow$ 45GBd  &       10.30       &  \textgreater 100     \\ \cline{2-4} 
                      & w/o TL           &          10.30    &    \textgreater 100   \\ \hline\hline
\multirow{3}{*}{TWC}  & TL 34.4GBd $\rightarrow$ 45GBd   &        9.50      &    \textless 10    \\ \cline{2-4} 
                      & TL 64GBd $\rightarrow$ 45GBd &      9.50        &  \textgreater 40     \\ \cline{2-4} 
                      & w/o TL           &          9.50    &   \textgreater 60    \\ \hline
\end{tabular}
\label{Table_TL_SR}
\end{table}

Finally, just as we did with the TL for the change in the launch power, we now explain why we use TL from the lower to the higher symbol rate scenarios. The studies we have done considering SSMF and TWC fiber are summarized in Table.~\ref{Table_TL_SR}. The analysis of Table.~\ref{Table_TL_SR} shows that transferring the learned features from a lower symbol rate to a higher one results in fewer epochs being required to train the NN-based equalizer, which indicates an increase in the TL effectiveness. This behavior is a consequence of reducing the impact of nonlinear effects that occur in optical fiber transmission by increasing the symbol rate while maintaining the same launch power. Just like when changing the launch power, we must once again choose the direction from a higher nonlinearity scenario to a lower one to improve the effectiveness of the TL.

\subsection{The operational limits of transfer learning for NN-based equalizers}\label{subSec:difffiber}

 The effectiveness of TL when multiple changes simultaneously occur in the transmission setup is discussed in this last subsection. 
 In this complicated case, we transfer the learning from the source and re-train all layers of the NN model, because now have the simultaneous change in the memory size and the intensity of nonlinearity.

To demonstrate the performance rendered by the TL, we address the following case. For SSMF, we transfer the learning from the source domain, i.e. from the setup operating at 8~dBm power, with 34.4~GBd symbol rate, with 16QAM (we can represent the domain parameters as \{8~dBm, 34.4~GBd, 16QAM\}), to \{6~dBm, 65~GBd, 64QAM\}; the second case SSMF studied is the transfer from  \{4~dBm, 34.4~GBd, 16QAM\}), to \{8~dBm, 65~GBd, 16QAM\}. For the TWC, we studied the TL from \{5~dBm, 34.4~GBd, 16QAM\} towards \{3~dBm, 65~GBd, 64QAM\}, and from \{8~dBm, 65~GBd, 16QAM\}), to \{4~dBm, 34.4~GBd, 16QAM\}. Our results are summarized in Table~\ref{Main_table}, rows 7 to 10. We can see that the TL is still useful even for such drastic scenario changes, and we can have variable but nonzero savings percentage in both the number of epochs and training dataset size, but the improvement can be much less pronounced as compared to the single parameter change scenarios. 

We note that when we increase the symbol rate while keeping the power and transmission setup the same, the power spectral density reduces. A reduced power spectral density implies a reduction in the nonlinear effects, and we have analyzed this in the previous subsection on the TL of symbol rates. However, in test cases 8 and 10 from Table~\ref{Main_table}, we address the cases where the source and the targets had approximately the same power spectral density because we change the symbol rate and the launch power proportionally. With this, we achieved some non-trivial results regarding the direction of the transfer. In test case 8, when using the TWC fiber which is characterized by a chromatic dispersion parameter that is 6 times smaller and by a nonlinear coefficient that is almost twice the one of SSMF, respectively, the best performance occurred when we followed the ``rule of power'' (i.e. going from a higher launch power to a lower one). However, in test case 10, where we considered the use of SSMF, we see that memory is the key factor. Thus, in this case, the best performance was achieved by following the ``rule of symbol rate'' (from low to high).

Finally, we study the possible performance limits of the TL technique and discuss the prospect of transferring the learned features between different fiber setups. For this analysis, we transfer the knowledge from the 18$\times$50~km SSMF link to the 9$\times$50~km TWC fiber link and vice versa, keeping the other transmission parameters the same: \{5~dBm, 34.4~GBd, 16QAM\}.
Table~\ref{Main_table}, rows 11 and 12, depicts the results for such a TL approach. We see that even when we change the fiber plant, and this is \textit{the largest change in channel function considered in our paper}, the TL can still provide a 50\% reduction in the re-training dataset size, thus demonstrating the potential of the TL approach. However, we were unable to identify a decrease in the number of necessary epochs for such a TL case. In fact, such a decrease in the TL effectiveness can be expected as the TL is well-tailored to different but still related source and target channel function \cite{pan2009survey,tan2018survey,lu2015transfer,zhuang2020comprehensive}. However, when the fiber type changes, the resulting channel posterior functions may become quite distinct, reducing the TL effectiveness.


\section{Conclusion}

In this paper, we investigated the applicability of TL methods for the adaptation and reconfiguration of NN-based equalizers in coherent optical systems. In particular, the potential of TL to reduce the number of training epochs and the training dataset without impacting the equalizer's performance was assessed. We demonstrated that transferring the knowledge is an effective solution for providing flexibility when using the NN-based impairments equalization. We have studied the peculiarities of transferring the learned features between different launch powers, modulation formats, symbol rates, and link setups. We believe that our findings are rather general, as we performed our tests considering two types of optical fibers, namely SSMF and TWC, and different scenarios; the similarity of the curves and tendencies in Figs. \ref{fig:transfer_power}--\ref{fig:transfer_SR} clearly underlines the universality of the TL method. Nevertheless, the potential of this technique still needs to be further validated experimentally. Importantly, we emphasize that the physical insight on the NN equalizer functioning can be used for the design of efficient TL techniques. Notably, when we transfer the learned features between different launch powers and modulation formats, we can retrain only the convolutional part of our equalizer, which brings about considerable savings in training complexity. In turn, when we transfer the learning between different symbol rates, which effectively translates into changing the memory size between the source and target systems, we can freeze the convolutional part and retrain only the biLSTM part. The latter, again, allows us to reduce the training complexity. The effectiveness of the transfer learning technique depends on the direction from which the learning is transferred. For the first time, we discuss the transfer direction when the system’s launch power, modulation format, and symbol rate are changed. Overall, we observed that we can reduce up to 99\% in terms of training epochs (required to achieve the best performance in the equalization) and 99\% in terms of training dataset size without affecting the performance quality of the retrained equalizer. We also addressed the most challenging cases corresponding to multiple changes in the system configuration, together with the swap of the fiber plant. Even for such a considerable scenario change, we have shown that the TL is still efficient, providing a reduction of up to 50\% of the training dataset size required to reach the optimal performance. The main findings of our research are summarized in Table \ref{Main_table}. Our results demonstrate that the TL is a powerful tool for the implementation of various NN-based equalization techniques in quickly changing real-world scenarios, providing the required flexibility, adaptability, and generalizability. 
\bibliographystyle{IEEEtran}
\bibliography{references}

\end{document}